\documentclass [10pt,twocolumn,oneside,a4paper] {article}
\usepackage{epsf}

\newcommand{\ec}{electron capture }
\newcommand{\bt}{$\beta$ }

\newcommand{\btm}{$\beta^{-}$ }
\newcommand{\gt}{Gamow-Teller }
\newcommand{\wi}{weak interaction }

\newcommand{\lt}{log(T) }

\def\be{\begin{equation}}
\def\ee{\end{equation}}
\def\bea{\begin{eqnarray}}
\def\eea{\end{eqnarray}}
\def\bdm{\begin{displaymath}}
\def\edm{\end{displaymath}}
\begin{document}
\onecolumn
\title {\textbf{Stellar Weak Interaction Rates and Energy Losses for $\mathbf{fp}$-Shell Nuclei Calculated in the Proton-Neutron Quasiparticle Random Phase Approximation (II).\\ (A = 61 to 80)} }
\author{\textbf{Jameel-Un Nabi}  \and
\textbf {Hans Volker Klapdor-Kleingrothaus} \\Max-Planck-Institut f\"ur Kernphysik, 69029 Heidelberg, Germany}
\normalsize
\maketitle
\begin{abstract}
Weak interaction rates and the associated energy losses for massive nuclei in the $fp$-shell are calculated in stellar matter using a modified form of proton-neutron quasiparticle RPA model with separable Gamow-Teller forces. A total of 209 nuclei with mass numbers ranging from A = 61 to 80 are considered here for the calculation of weak rates. These also include many neutron-rich nuclei which play a key role in the dynamics of the core collapse calculations. The stellar weak rates are calculated over a wide range of densities (10$\leq$ $\rho Y_{e}$ (gcm$^{-3}$) $ \leq$ 10$^{11}$) and temperatures (10$^{7}$ $\leq$ T(K) $\leq$ 30 $\times$ 10$^{9}$). This is our third paper in a series starting with the calculation of stellar weak rates for nuclei in the $sd$-shell [ATOMIC DATA AND NUCLEAR DATA TABLES \textbf{71}, 149 (1999)]. The calculated rates take into consideration the latest experimental energy levels and $ft$ value compilations. The effect of particle emission processes is taken into account and the energies and probabilities of these particle emission processes are also calculated in stellar environment. Our results are presented here on an abbreviated scale of temperature and density.
\end{abstract}
\clearpage
\twocolumn
\section {INTRODUCTION}
Gravitational collapse of the core of massive stars ($M > 10M_{\odot}$) is triggered by electron capture on free protons and on nuclei at high temperatures and densities. These stars then end up either as a neutron star or a black hole by supernova explosion (Type II). Electron capture during the late stages of stellar evolution was discussed in detail by \cite{Bet79}. The weak interaction certainly plays a key role in the dynamics of the core collapse. Hundreds, even thousands of nuclear reactions immediately before and after the collapse are relevant, and many of these concern unstable nuclei that are difficult to study experimentally. Therefore detailed, reliable theory is needed in order to understand these violent processes. 

Because the final outcome of the explosion depends so sensitively on a variety of physical inputs at the beginning of each stage of the entire process (i.e., collapse, shock formation, and shock propagation), it is desirable to calculate the presupernova stellar structure with the best possible physical data and input currently available. The energy budget would be balanced in favor of an explosion by a smaller precollapse iron core mass and a lower entropy. Since the hydrodynamic shock, believed to be responsible for the explosion, forms at the edge of the homologous core (the inner part of the core which remains in acoustic communication throughout core collapse and bounce), a smaller iron core size means that the energy lost by the shock in photodisintegrating iron nuclei in the overlying matter would be correspondingly smaller \cite{Bur83}. As less energy is stored in nuclear excited states in the lower entropy environment of the presupernova as well as the collapsing core, the collapse can proceed to a higher density and generate a stronger bounce and form a more energetic shock wave \cite{Bet79}. In addition, at smaller entropy the abundance of free protons remains less. Since protons are the main sink for electrons through the electron capture reactions, a smaller initial entropy even at the presupernova stage means a lower overall electron capture rate and therefore a higher value of $Y_{e}$ (the number of electrons per baryon) at the time of bounce. This in turn makes the mass $M_{HC}$ of the unshocked  inner core (the homologous core) larger. As a result, the shock energy
\begin{eqnarray*}
E_{S} \simeq (G M_{HC}^{2}/R_{HC})(Y_{ef}-Y_{ei}) \simeq M_{HC}^{5/3}(Y_{ef}-Y_{ei}) \nonumber\\
 \simeq Y_{ef}^{10/3}(Y_{ef}-Y_{ei}),
\end{eqnarray*}
is larger for larger final lepton fraction $Y_{ef}$, where $R_{HC}$ is the homologous core radius and $Y_{ei}$ is the initial electron fraction.

The nuclear weak-interaction mediated reactions, e.g., \bt decay and \ec, not only lead to a change in the neutron-to-proton ratio in the stellar core material but because of the removal of the energy by neutrinos produced in the reactions, they cool the core to a lower entropy state. It is therefore important to follow the evolution of the stellar core during its late stages of hydrostatic nuclear burning with a sufficiently detailed nuclear reaction network that includes these weak-interaction mediated reactions.

In the aftermath of the collapse, r-process nucleosynthesis (theory of which was put forward by \cite{Bur57,Cam57}) hinges on the waiting-point nuclides, where the \bt decay rate is long enough for substantial amounts of material to build up. This leads to the observed r-process peaks in isotopic abundances \cite{Arn96}.

As already mentioned there are a huge number of nuclear reactions that go into modeling a supernova and its aftermath, from the formation of the presupernova core, to the high neutron flux and rapid capture of neutrons after the collapse. The nuclear physics is challenging, in part because of the sheer number of reactions as well as the empirical inaccessibility of some of the unstable intermediate nuclides. The challenges to theory include modeling medium mass nuclei, typically A $\sim$ 50--90 and at the same time handling of a large number of excited states involved. The highly-anticipated experiments with unstable nuclei at radioactive ion beam accelerator facilities under construction and being upgraded around the world \cite{RIB} will strongly test and constrain the models. 

The main task is to find a model with an appropriate number of degrees of freedom. Any model with too few microscopic degrees of freedom would be inadequate; conversely, a model with too many degrees of freedom (such as full-blown shell model calculations, which for iron-region nuclides could easily require on the order of 10$^{9}$ basis states) could be computationally intractable.

Fuller, Fowler and Newman \cite{Ful82} (hereafter refered to as FFN) did the most extensive calculations of stellar weak rates over a wide range of temperature (10$^{7}$ $\leq$ T(K) $\leq$ 10$^{11}$) and density (10 $\leq$ $\mathit{\rho Y_{e}}$ (gcm$^{-3}$) $\leq$ 10$^{11}$). They calculated stellar electron and positron emission rates and continuum electron and positron capture rates, as well as the associated neutrino energy loss rates for 226 nuclei with masses between A = 21 and 60. Measured nuclear level information and matrix elements available at that time were used and unmeasured matrix elements for allowed transitions were assigned an average value of $log ft = $5. To complete the FFN rate estimate, the Gamow-Teller contribution to the rate was parameterized on the basis of the independent particle model and supplemented by a contribution simulating low-lying transitions. Soon it was realized that calculations of presupernova evolution generate cores that are so neutron-rich ($Y_{e} \leq 0.42$) that nuclei more massive than A = 60 must be considered \cite{Auf90}. The FFN rates were then updated and extended to heavier nuclei \cite{Auf94}.
 Beta decays previously were thought not to play an important role in the process of the core collapse and were not included in evolutionary calculations (see for example \cite{Wea78,Wea85}). It was shown in the work of \cite{Auf94} that beta decays indeed play an important role in the evolution of the core below $Y_{e} \sim 0.46 $ as the core approaches a state of dynamic equilibrium between electron captures and beta decays. Motivated by the ideas of these papers \cite{Auf94,Auf90} Kar et al. \cite{Kar94} calculated \bt decay rates for 11 nuclei in the mass range A $>$ 60 using an average beta strength function for typical presupernova matter density ($\rho = 3 \times 10^{7}$ to $3 \times 10^{9}$ g ~cm$^{-3}$) and temperature ($T = (2$ to $5) \times 10^{9}$K).  

First microscopic calculations \cite{Kla84,Sta90a,Sta89,Hir93} of the terrestrial rates used the pn-QRPA theory, and were found to give good agreement with the experimental data. The pn-QRPA theory is a good compromise for the calculation of \wi rates in stellar environment between the phenomenological calculations described above and the full-blown shell model calculations where the number of basis states is too big to handle. The shell model calculations can be done on price of truncation of the model space but then there is a chance of missing certain nuclear effects. An alternate method of solution, Monte Carlo evaluation of path integrals, also known as Monte Carlo shell model \cite{Joh92}, allows access to larger model spaces but at the cost of restricted spectroscopy and a limited class of interactions.
  
This work is the first ever extensive calculation of stellar weak rates in the $fp$- and $fpg$-shell nuclei ranging from A = 40 to 100 using the pn-QRPA theory (see also \cite{Nab99a}). A total of 619 nuclei were considered for the calculation of stellar weak rates. These include also proton-rich and neutron-rich nuclei. This work is a continuation of our first paper for $fp$-shell nuclei where twelve different weak rates had been calculated for each parent nucleus. These included $e^{\pm}$-capture rates, $\beta^{\pm}$-decay rates, neutrino (anti-neutrino) energy loss rates, energies of beta-delayed proton (neutron) and the probabilities of these beta-delayed particle emission processes.  

Section~2 reviews briefly our formalism of calculations which we think is necessary to restate. This can help in independent study of this paper. For details of formalism we refer to \cite{Nab99}. Results, discussions and comparisons with previous calculations are dealt with in Section~3. We finally summarize our conclusions in Section~4. 

\section{FORMALISM}
The following main assumptions are made:

(1) Allowed \gt and superallowed Fermi transitions were taken into account for our calculations. Contributions from forbidden transitions were assumed to be negligible.

(2) The atoms are totally ionized as a result of the high temperature prevailing in the stellar interior, and the electrons obey the energy distribution function of a Fermi gas. At high temperatures (kT $>$ 1~MeV), positrons appear via electron-positron pair creation, and the positrons follow the same energy distribution function as the electrons.
 
(3) Neutrinos and antineutrinos escape freely from the interior of the star. Therefore, there are no (anti)neutrinos which block the emission of these particles in the capture or decay processes. Also, (anti)neutrino capture is not taken into account.

(4) The distortion of electron (positron) wavefunction due to the Coulomb interaction with a nucleus is represented by the Fermi function in phase space integrals.

(5) We consider the effect of particle emission processes from excited states. All excited states, with energy greater than $S_{p}$ or $S_{n}$, are assumed to decay to the ground state by emission of protons or neutrons, respectively. Due to uncertainties in the calculation of energy levels and the effect of the Coulomb barrier for the case of proton emission, it is assumed that particles are emitted at excited energies 1~MeV higher (called $E_{crit}$ throughout this section) than the particle decay channel. For the parent nuclei, the cut-off excitation energy is set at $E_{crit}$. All excited states in daughter nuclei, with excitation energy ($E_{j} \leq E_{crit}$) decay directly to the ground state through $\gamma$ transitions. All excited energy states ($E_{j} > E_{crit}$) decay by emitting particles (protons or neutrons) with increasing kinetic energies. It is further assumed that either protons (if $S_{p} < S_{n}$) or neutrons (if $S_{n} < S_{p}$) are emitted from these excited states. The $\beta$ decay of a possible isomer is not taken into account.

\subsection{Rate Formulae}
The decay (capture) rate of a transition from the $i$th state of a parent nucleus ($Z, N$) to the $j$th state of the daughter nucleus ($Z \mp 1, N \pm 1$) is given by  \footnote{Throughout subsection 2.1 natural units $(\hbar=c=m_{e}=1)$ are adopted, unless otherwise stated, where $m_{e}$ is the electron mass.}
\begin{equation}
\lambda_{ij} =ln2 \frac{f_{ij}(T,\rho,E_{f})}{(ft)_{ij}},
% = \frac{ln2}{D}B_{ij}g_{ij}
\end{equation}
where $(ft)_{ij}$ is related to the reduced transition probability $B_{ij}$ of the nuclear transition by
\be
(ft)_{ij}=D/B_{ij}.
\ee
The $D$ appearing in Eq.~(2) is a compound expression of physical constants,
\be
D=\frac{2ln2\hbar^{7}\pi^{3}}{g_{V}^{2}m_{e}^{5}c^{4}},
\ee
and,
\be
B_{ij}=B(F)_{ij}+(g_{A}/g_{V})^2 B(GT)_{ij},
\ee
where B(F) and B(GT) are reduced transition probabilities of the Fermi and ~Gamow-Teller transitions respectively,
\be
B(F)_{ij} = \frac{1}{2J_{i}+1} \mid<j \parallel \sum_{k}t_{\pm}^{k} \parallel i> \mid ^{2},
\ee 
\be
B(GT)_{ij} = \frac{1}{2J_{i}+1} \mid <j \parallel \sum_{k}t_{\pm}^{k}\vec{\sigma}^{k} \parallel i> \mid ^{2},
\ee

In Eq.~(6), $\vec{\sigma}^{k}$ is the spin operator and $t_{\pm}^{k}$ stands for the isospin raising and lowering operator. The value of D=6295 s is adopted and the ratio of the axial-vector $(g_{A})$ to the vector $(g_{V})$ coupling constant is taken as 1.254. 

The phase space integral $(f_{ij})$ is an integral over total energy,
\be
f_{ij} = \int_{1}^{w_{m}} w \sqrt{w^{2}-1} (w_{m}-w)^{2} F(\pm Z,w) (1-G_{\mp}) dw,
\ee
for electron (\textit{upper signs}) or positron (\textit{lower signs}) emission, or 
\be
f_{ij} = \int_{w_{l}}^{\infty} w \sqrt{w^{2}-1} (w_{m}+w)^{2} F(\pm Z,w) G_{\mp} dw,
\ee
for continuum positron (\textit{lower signs}) or electron (\textit{upper signs}) capture.

In Eqs.~(7) and (8), $w$ is the total kinetic energy of the electron including its rest mass, $w_{l}$ is the total capture threshold energy (rest+kinetic) for positron (or electron) capture. One should note that if the corresponding electron (or positron) emission total energy, $w_{m}$, is greater than -1, then $w_{l}=1$, and if it is less than or equal to 1, then $w_{l}=\mid w_{m} \mid$. $w_{m}$ is the total $\beta$-decay energy,
\be
w_{m} = m_{p}-m_{d}+E_{i}-E_{j},
\ee
where $m_{p}$ and $E_{i}$ are mass and excitation energies of the parent nucleus, and $m_{d}$ and $E_{j}$ of the daughter nucleus, respectively.

$G_{+}$ and $G_{-}$ are the positron and electron distribution functions, respectively. Assuming that the electrons are not in a bound state, these are the Fermi-Dirac distribution functions,
\be
G_{-} = [exp (\frac{E-E_{f}}{kT})+1]^{-1},
\ee
\be
G_{+} = [exp (\frac{E+2+E_{f}}{kT})+1]^{-1}.
\ee
Here $E=(w-1)$ is the kinetic energy of the electrons, $E_{f}$ is the Fermi energy of the electrons, $T$ is the temperature, and $k$ is the Boltzmann constant.

The number density of electrons associated with protons and nuclei is $\rho Y_{e} N_{A}$, where $\rho$ is the baryon density, $Y_{e}$ is the ratio of electron number to the baryon number, and $N_{A}$ is the Avogadro number.
\be
\rho Y_{e} = \frac{1}{\pi^{2}N_{A}}(\frac {m_{e}c}{\hbar})^{3} \int_{0}^{\infty} (G_{-}-G_{+}) p^{2}dp, 
\ee
where $p=(w^{2}-1)^{1/2}$ is the electron or positron momentum, and Eq.~(12) has the units of \textit{moles $cm^{-3}$}. This equation is used for an iterative calculation of Fermi energies for selected values of $\rho Y_{e}$ and $T$.

In the calculations, the inhibition of the final neutrino phase space is never appreciable enough that neutrino (or anti-neutrino) distribution functions had to be taken into consideration. $F(\pm Z,w)$ are the Fermi functions and are calculated according to the procedure adopted by Gove and Martin \cite{Gov71}.

There is a finite probability of occupation of parent excited states in the stellar environment as a result of the high temperature in the interior of massive stars. Weak interaction rates then also have a finite contribution from these excited states. The occupation probability of a state $i$ is calculated on the assumption of thermal equilibrium,
\be
P_{i} = \frac {(2J_{i}+1)exp(-E_{i}/kT)}{\sum_{i=1}(2J_{i}+1)exp(-E_{i}/kT)},
\ee
where $J_{i}$ and $E_{i}$ are the angular momentum and excitation energy of the state $i$, respectively.

Unfortunately one cannot calculate the $J_{i}$'s in QRPA theory and hence Eq.~(13) is modified as
\be
P_{i} = \frac {exp(-E_{i}/kT)}{\sum_{i=1}exp(-E_{i}/kT)}.
\ee
This approximation is a compromise and can be justified when one takes into consideration the uncertainty in the calculation of $E_{i}$ which easily over-sheds the uncertainty in calculating the values of $J_{i}$ in the above Eq.~(13).

The rate per unit time per nucleus for any weak process is then given by
\be
\lambda = \sum_{ij}P_{i} \lambda_{ij}.
\ee
The summation over all initial and final states is carried out until satisfactory convergence in the rate calculations is achieved.

The neutrino energy loss rates are calculated using the same formalism except that the phase space integral is replaced by
\be
f_{ij}^{\nu} = \int_{1}^{w_{m}} w \sqrt{w^{2}-1} (w_{m}-w)^{3} F(\pm Z,w) (1-G_{\mp}) dw,
\ee
for electron (\textit{upper signs}) or positron (\textit{lower signs}) emission, or by
\be
f_{ij}^{\nu} = \int_{w_{l}}^{\infty} w \sqrt{w^{2}-1} (w_{m}+w)^{3} F(\pm ,w) G_{\mp} dw,
\ee
for continuum positron (\textit{lower signs}) or electron (\textit{upper signs}) capture.

 The $\lambda_{ij}$ in Eq.~(15) in this case is the sum of the \ec and positron decay rates, or the sum of the positron capture and electron decay rates, for the transition $i \rightarrow j$.

The proton energy rate from the daughter nucleus is calculated, whenever $S_{p}$ $<$ $S_{n}$, by
\be
\lambda^{p} = \sum_{ij}P_{i}\lambda_{ij}(E_{j}-E_{crit}),
\ee
for all $E_{j} > E_{crit}$, whereas for all $E_{j} \leq E_{crit}$ one calculates the $\gamma$ heating rate,
\be
\lambda^{\gamma} = \sum_{ij}P_{i}\lambda_{ij}E_{j}.
\ee
If on the other hand, $S_{n} < S_{p}$, then the neutron energy rate from the daughter nucleus is calculated using
\be
\lambda^{n} = \sum_{ij}P_{i}\lambda_{ij}(E_{j}-E_{crit}),
\ee
for all $E_{j} > E_{crit}$, and for all $E_{j} \leq E_{crit}$ the $\gamma$ heating rate is calculated as in Eq.~(19).  

The probability of $\beta$-delayed proton (neutron) emission is calculated by
\be
P^{p(n)} = \frac{\sum_{ij\prime}P_{i}\lambda_{ij\prime}}{\sum_{ij}P_{i}\lambda_{ij}},
\ee
where $j\prime$ are states in the daughter nucleus for which $E_{j\prime} > E_{crit}$. In Eqs.~[(18)-(21)] $\lambda_{ij(\prime)}$ is the sum of the electron capture and positron decay rates, or the sum of positron capture and electron decay rates, for the transition $i$ $\rightarrow$ $j(j\prime)$. For a detailed formalism of the calculation of excited states and the \gt transition amplitudes between these states as well as the Fermi transition amplitudes we refer to \cite{Nab99}.

\subsection{Incorporation of Experimental Data}
The latest  experimental excitation energies and log $\mathit{ft}$ values are incorporated in our calculations wherever possible to increase the reliability of our results. 

The calculated excitation energies were replaced with the measured ones when they were within 0.5 MeV of each other. The log $\mathit{ft}$ value of this energy level was also then replaced by the measured one. Very low lying states were inserted in the calculations together with their log $\mathit{ft}$ values if the theory was missing them. Inverse and mirror transitions were also taken into consideration. If there appeared a level in experimental compilations without definite spin and parity assignment, no theoretical levels were replaced with experimental levels beyond this excitation energy nor were they inserted in the calculations.

The Q-value of each transition as well as $S_{p}$ and $S_{n}$ of each nucleus were derived using the experimental mass compilation of Audi et al. \cite{Aud95}. For nuclei where \cite{Aud95} failed to give mass defects predictions of M\"oller and Nix \cite{Moe81} and Myers and Swiatecki \cite{Mye96} were used to derive the corresponding energies necessary for the calculations. The two different theoretical mass formulae used reflect the sensitivity of stellar rates on input masses. A total of 31 out of the 209 nuclei (83 out of the total 619 for all $fp$- and $fpg$-shell nuclei considered in our calculations) presented here were calculated using theoretical mass formulae. Stellar rates using mass formula of Myers and Swiatecki \cite{Mye96}, for the case of neutron-rich and proton-rich nuclei, are given in this paper on an abbreviated scale of density and marked accordingly (Table~B). In \cite{Nab99t} the stellar rates of neutron-rich nuclei using the mass model of M\"oller and Nix \cite{Moe81} are presented. The complete rate table on magnetic tapes can be requested from the authors.

Table~(1) shows the references from which the latest experimental excitation energies and log $\mathit{ft}$ values are taken (NP $\rightarrow$ Nuclear Physics and NDS $\rightarrow$ Nuclear Data Sheets). For those nuclei where more than one references were given, the latest data was taken into account. 
\section{RESULTS AND DISCUSSION}
As stated earlier there are not many calculations available for stellar weak rates for nuclei A $>$ 60. A compilation of beta-decay rates for 11 nuclei in the mass range A $>$ 60 was done by Kar et al. \cite{Kar94} (which we will refer to as KRS in this paper). Some assessment of \ec and \bt decay rates for these heavy nuclei was also reported by Aufderheide et al. \cite{Auf90,Auf94} (which we will refer to as ABKSV and AFWH, respectively, in this paper). Transitions from only the lowest state of the parent nucleus having an allowed transition were calculated in a very crude way by ABKSV. This was also admitted by the authors and comparison of these rates with the QRPA calculations clearly reflects the situation (see section 3.2). At that time no microscopic calculations for $fp$-shell nuclei were available. In their later calculations of \ec and \bt decay, AFWH estimated the \bt decay rates in a similar fashion as the \ec rates, and phenomenologically parameterized the position and strength of the back resonance. This estimate was supplemented by an empirical contribution, placed at zero excitation energy, which simulates low-lying strength missed by the \gt resonance. Below a comparison of these earlier compilations with this work is presented.
\subsection{Comparison with the calculation of KRS.}
KRS constructed a model to calculate the beta-decay rates using an average beta strength function and an electron-phase space factor evaluated for typical presupernova matter density ($\rho$ = 3 $\times$ 10$^{7}$ to 3 $\times$ 10$^{9}$ g cm$^{-3}$) and temperature ($T$ = 2 $\times$ 10$^{9}$ to 5 $\times$ 10$^{9}$ K). For the \gt strength function KRS used  a sum rule calculated by the spectral distribution theory, and the centroid of the distribution was obtained from the experimental data on (p,n) reactions. The width, $\sigma$, of the \gt strength function had two parts ($\sigma^{2} = \sigma^{2}_{N} + \sigma^{2}_{C}$, with $\sigma_{C}$ = 0.157 Z A$^{-1/3}$). The parameter $\sigma_{N}$ was fixed by a best fit to the observed half-lives for the free decays of a number of A $>$ 60 nuclei to a global average value of 6.3 MeV.
\begin{figure}
\epsfxsize=7.8cm
\epsffile{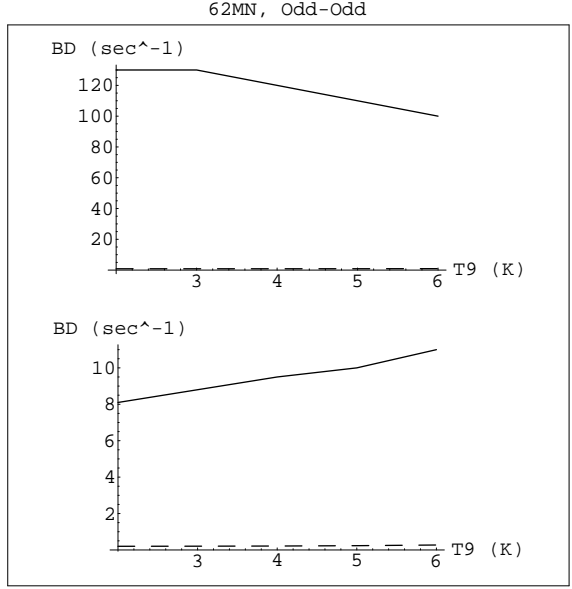}
\caption{ \footnotesize Comparison of the QRPA  beta-decay (BD) rates (this work) with those of [12]. Solid lines represent the QRPA beta-decay rates while broken lines represent the decay rates of [12]. Log(T) is the log of temperature in units of Kelvin and BD represents the \btm rates in units of sec$^{-1}$. The upper graph is plotted at $\rho Y_{e}=$ 1.6 $\times$10$^{7}$ g cm$^{-3}$ and the lower graph is plotted at $\rho Y_{e}=$ 1.6 $\times$10$^{9}$ g cm$^{-3}$.}
\end{figure}
\begin{figure}
\epsfxsize=7.8cm
\epsffile{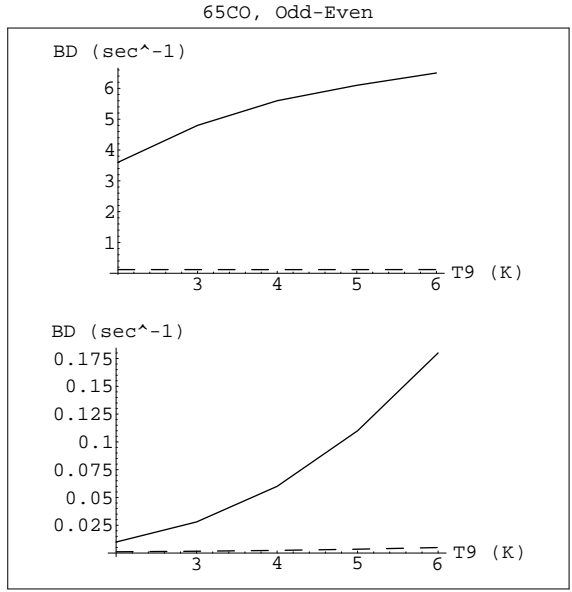}
\caption{ \footnotesize Same as Fig.~1 but for \bt decay of $^{65}$Co.}
\end{figure}

Fig.~1 shows the comparison of beta-decay rates for $^{62}$Mn. One sees that at the lower density and \lt = 2, the QRPA rate (calculated in this work) is enhanced by more than an order of magnitude. KRS considered just the transitions from the ground state of $^{62}$Mn whereas the QRPA rates considered $E_{i}$ up to 5.802~MeV. This is the reason that the rates of KRS are so much suppressed. At high temperatures there is a finite probability of occupation of parent excited states and KRS did not consider the transitions from these excited states. At higher density ($\rho Y_{e}=$ 1.6 $\times$10$^{9}$ g cm$^{-3}$) the respective rates are further suppressed. For nuclei with low or intermediate Q-values, the beta-decay rates tend to go down at high densities due to the blocking of the available phase space of increasingly degenerate electrons. At high densities, the contribution to the rates by the excited states is very important, especially for the higher temperatures, because such states, due to their higher Q-values, are still unblocked. In contrast, at lower densities, the differences between the ground-state transition rates and the rates inclusive of the excited states are not equally large.
\begin{figure}
\epsfxsize=7.8cm
\epsffile{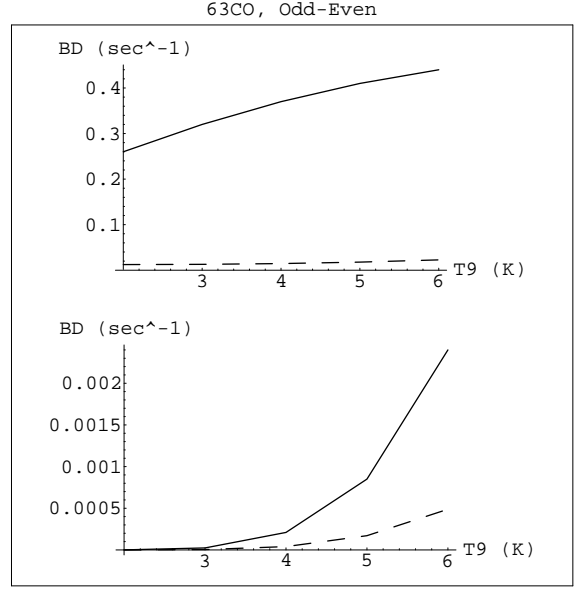}
\caption{ \footnotesize Similar to Fig.~1. The upper graph is plotted at $\rho Y_{e}=$ 1.3 $\times$10$^{7}$ g cm$^{-3}$ and the lower graph is plotted at $\rho Y_{e}=$ 1.3 $\times$10$^{9}$ g cm$^{-3}$ for \bt decay of $^{63}$Co.}
\end{figure}   
\begin{figure}
\epsfxsize=7.8cm
\epsffile{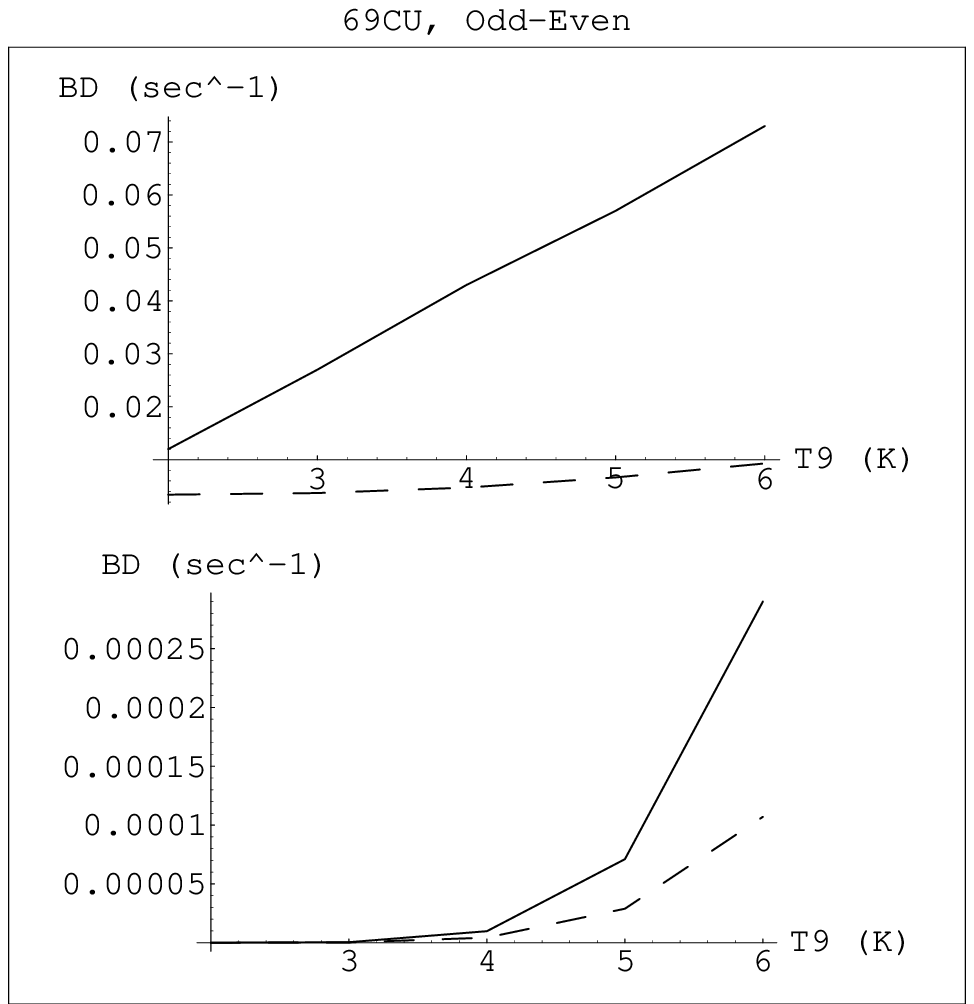}
\caption{ \footnotesize Same as Fig.~3 but for \bt decay of $^{69}$Cu.}
\end{figure}
\begin{figure}
\epsfxsize=7.8cm
\epsffile{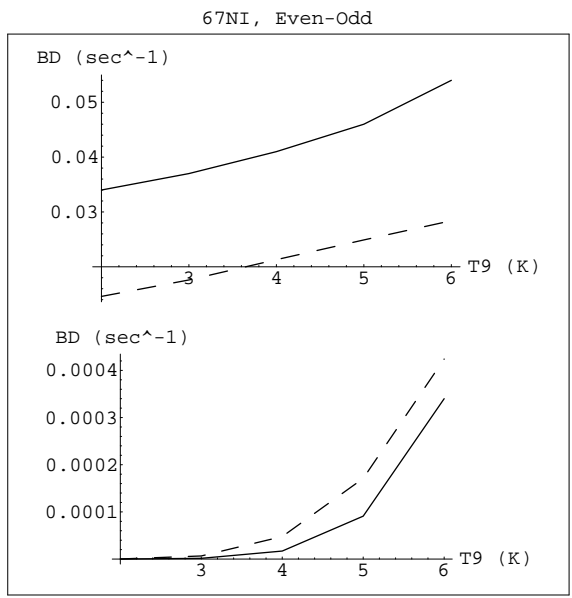}
\caption{ \footnotesize Same as Fig.~3 but for \bt decay of $^{67}$Ni.}
\end{figure}
Figures~2--4 show similar comparisons but one notes that the rates of KRS get in better and better comparison with the QRPA rates. In Fig.~5, eg., for the case of beta-decay of $^{67}$Ni, the QRPA rate is just about a factor 2 greater than that of KRS decay rate (at \lt = 2.0 and  $\rho Y_{e}=$ 1.3 $\times$10$^{7}$ g cm$^{-3}$). At \lt = 2.0 and  $\rho Y_{e}=$ 1.3 $\times$10$^{9}$ g cm$^{-3}$ the comparison is again fairly good. In case of $^{67}$Ni, KRS went up to $E_{i}$ of 3.680 MeV as compared to 6.785 MeV considered by the QRPA rates.
\begin{figure}
\epsfxsize=7.8cm
\epsffile{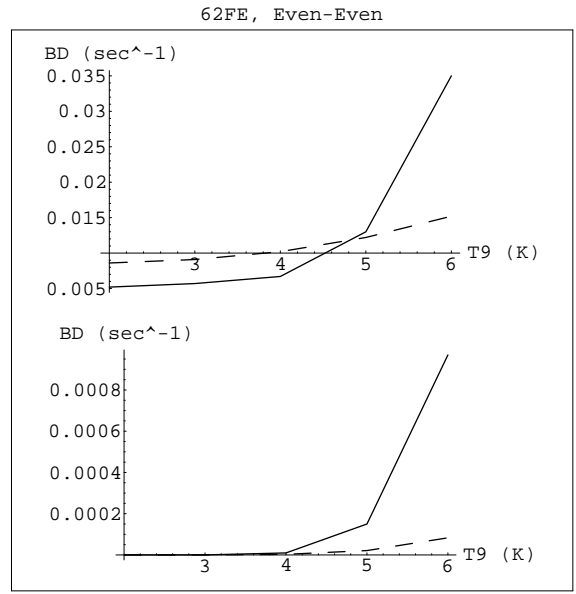}
\caption{ \footnotesize Same as Fig.~1 but for \bt decay of $^{62}$Fe.}
\end{figure}
\begin{figure}
\epsfxsize=7.8cm
\epsffile{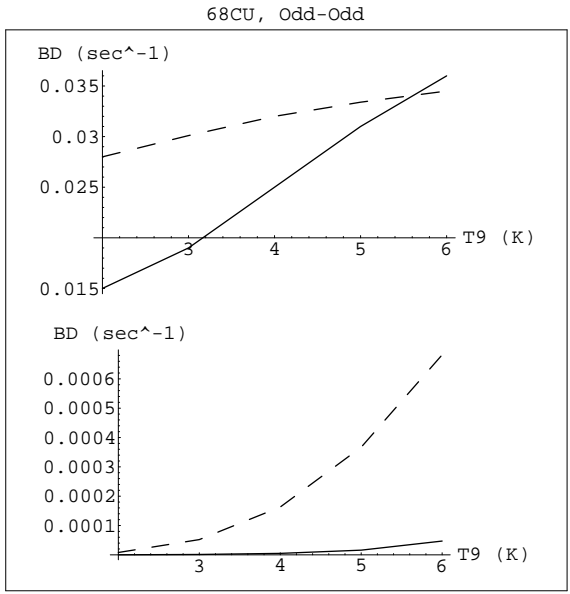}
\caption{ \footnotesize Same as Fig.~3 but for \bt decay of $^{68}$Cu.}
\end{figure}
As previously described, the parameter $\sigma_{N}$, in KRS, was fixed by a best fit to the observed half-lives for the free decays of a number of A $>$ 60 nuclei to a global average value of 6.3 MeV. For certain cases, KRS rates are also enhanced in comparison to the QRPA rates. This seems to be the case for $^{62}$Fe and $^{68}$Cu. Fig.~6 and Fig.~7 depict these cases, respectively. In both figures one notes that at lower density the beta-decay rates of KRS exceed the QRPA rates. At \lt = 2.0 the QRPA rate is suppressed by a factor of 1.7 for the case of $^{62}$Fe and by a factor of 1.9 for $^{68}$Cu. However at higher temperatures and densities, again, the QRPA rates surpass the beta-decay rates of KRS as the authors take a few parent excited states into consideration.
\begin{figure}
\epsfxsize=7.8cm
\epsffile{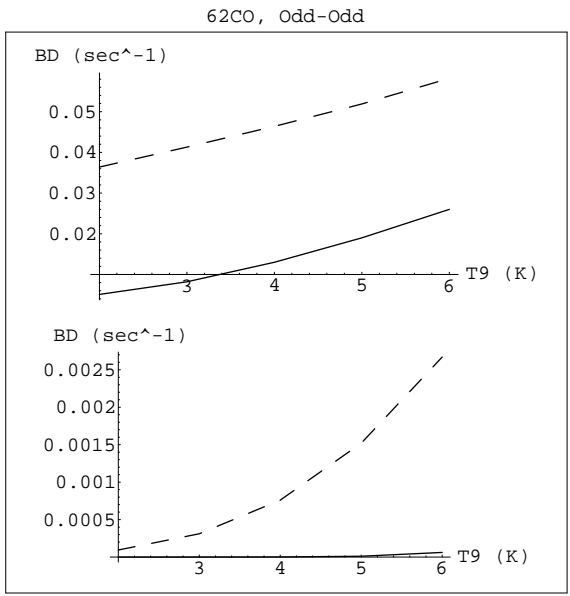}
\caption{ \footnotesize Same as Fig.~1 but for \bt decay of $^{62}$Co.}
\end{figure}
\begin{figure}
\epsfxsize=7.8cm
\epsffile{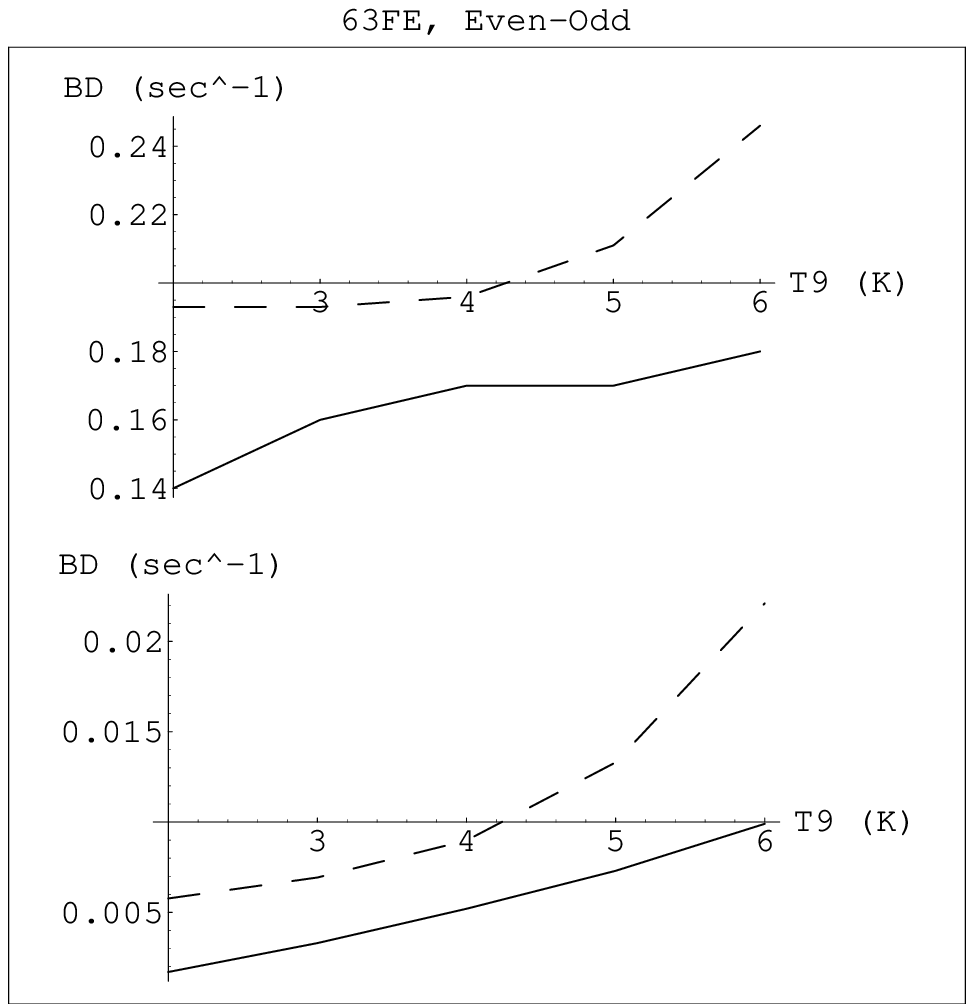}
\caption{ \footnotesize Same as Fig.~1 but for \bt decay of $^{63}$Fe.}
\end{figure}
\begin{figure}
\epsfxsize=7.8cm
\epsffile{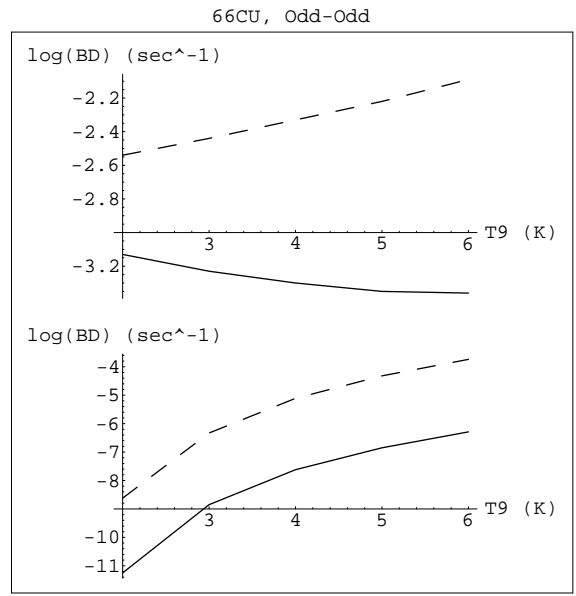}
\caption{ \footnotesize Same as Fig.~3 but for \bt decay of $^{66}$Cu.}
\end{figure} 
Fig.~8, Fig.~9 and Fig.~10 are the cases where the rates of KRS  are enhanced at all points of temperature and density. In Fig.~10 the abscissa represents the log of beta-decay rates so in this extreme case the beta-decay rates of KRS are enhanced by a factor of 3.9 at \lt = 2.0 and $\rho Y_{e}=$ 1.3~$\times$~10$^{7}$~g~cm$^{-3}$ and by a factor of 427 at \lt = 2.0 and $\rho Y_{e}=$1.3 $\times $10$^{9}$ gcm$^{-3}$.
\begin{figure}
\epsfxsize=7.8cm
\epsffile{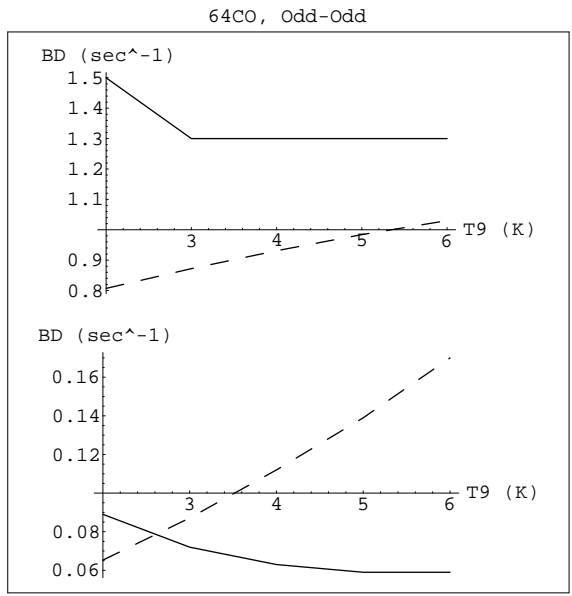}
\caption{ \footnotesize Same as Fig.~1 but for \bt decay of $^{64}$Co.}
\end{figure}
Fig.~11 represents the special case of $^{64}$Co. In this case the strength function of KRS over-estimated the terrestrial half-life of $^{64}$Co by an order of magnitude \cite{Kar94}. They were then forced to change the shape of the average beta-strength function. For this case the authors changed the value of $\sigma_{N}$ from the global value of 6.3 MeV to 7.5 MeV and also distorted the Gaussian of the \gt strength distribution introducing a negative skewness in the distribution to bring more strength in the ground-state domain. One sees that at lower density their  \gt strength function is still not giving strong transitions and the QRPA rate is enhanced by a factor of 1.9 at \lt = 2.0 and $\rho Y_{e}=$ 1.6 $\times$10$^{7}$ g cm$^{-3}$. Only at $\rho Y_{e}=$ 1.6 $\times$10$^{9}$ g cm$^{-3}$ their rates start overtaking the QRPA rates at \lt $>$ 3.0.  
\subsection{Comparison with the calculations of ABKSV and AFWH.}
The \ec rates of certain isotopes of copper and cobalt were calculated considering transitions from only the lowest state of the parent nucleus having an allowed transition \cite{Auf90}. The authors used many approximations in calculating the \gt strength function. A comparison of all the \ec rates of nuclei of A $>$ 60 (from ABKSV) is presented below.
\begin{figure}
\epsfxsize=7.8cm
\epsffile{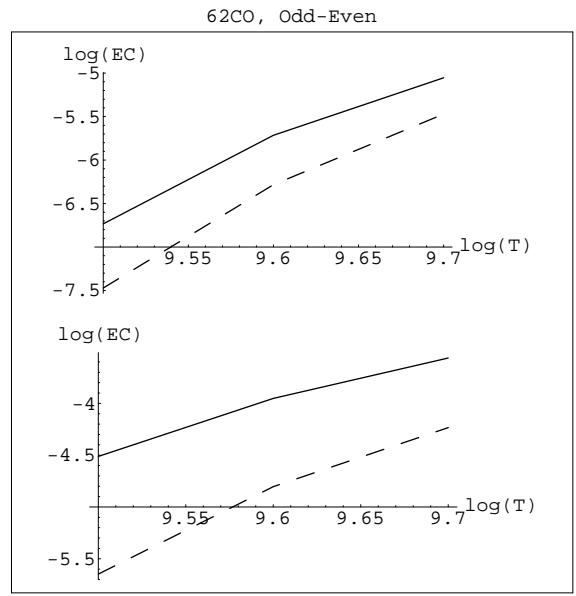}
\caption{ \footnotesize Comparison of the QRPA  \ec (EC) rates (this work) with those of [8]. Solid lines represent the QRPA \ec rates while broken lines represent the \ec rates of [8]. Log(T) is the log of temperature in units of Kelvin and log(EC) represents the log of \ec rates in units of sec$^{-1}$. The upper graph is plotted at $\rho Y_{e}=$ 10$^{7}$ g cm$^{-3}$ and the lower graph is plotted at $\rho Y_{e}=$ 10$^{8}$ g cm$^{-3}$.}
\end{figure}
\begin{figure}
\epsfxsize=7.8cm
\epsffile{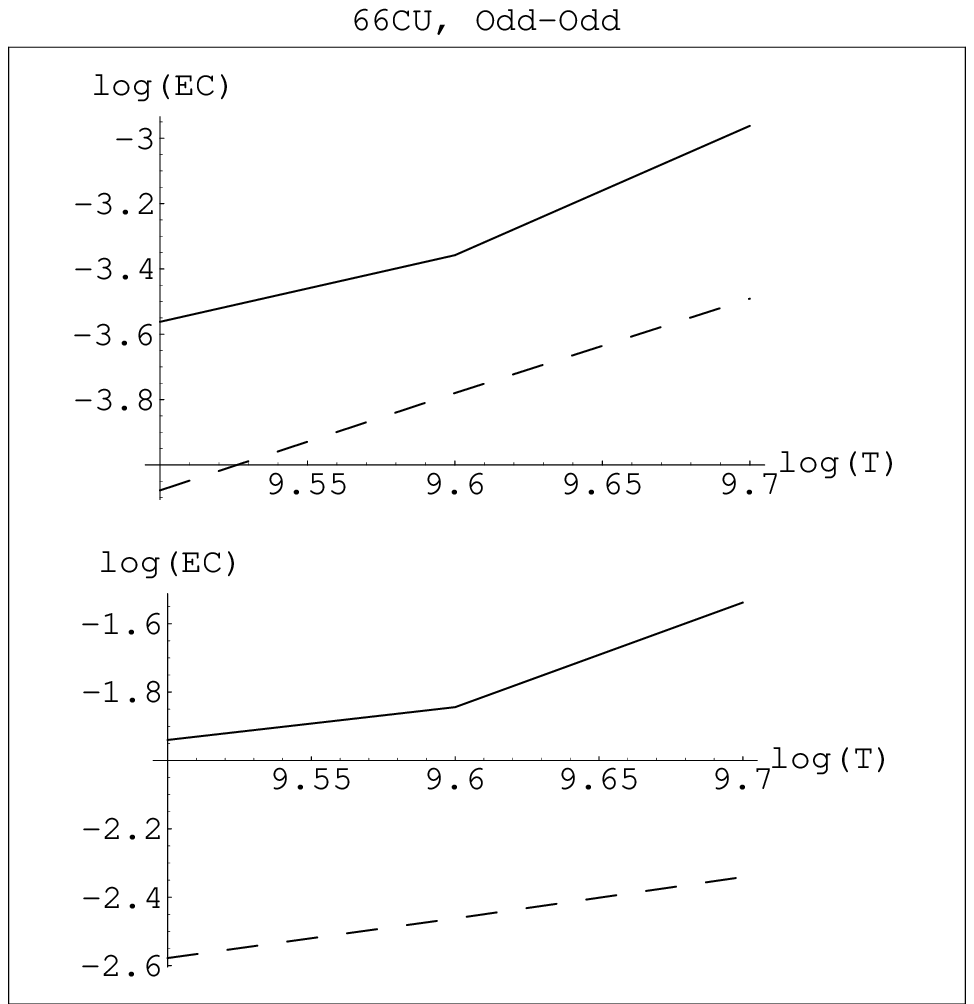}
\caption{ \footnotesize Same as Fig.~12 but for \ec by $^{66}$Cu.}
\end{figure}
\begin{figure}
\epsfxsize=7.8cm
\epsffile{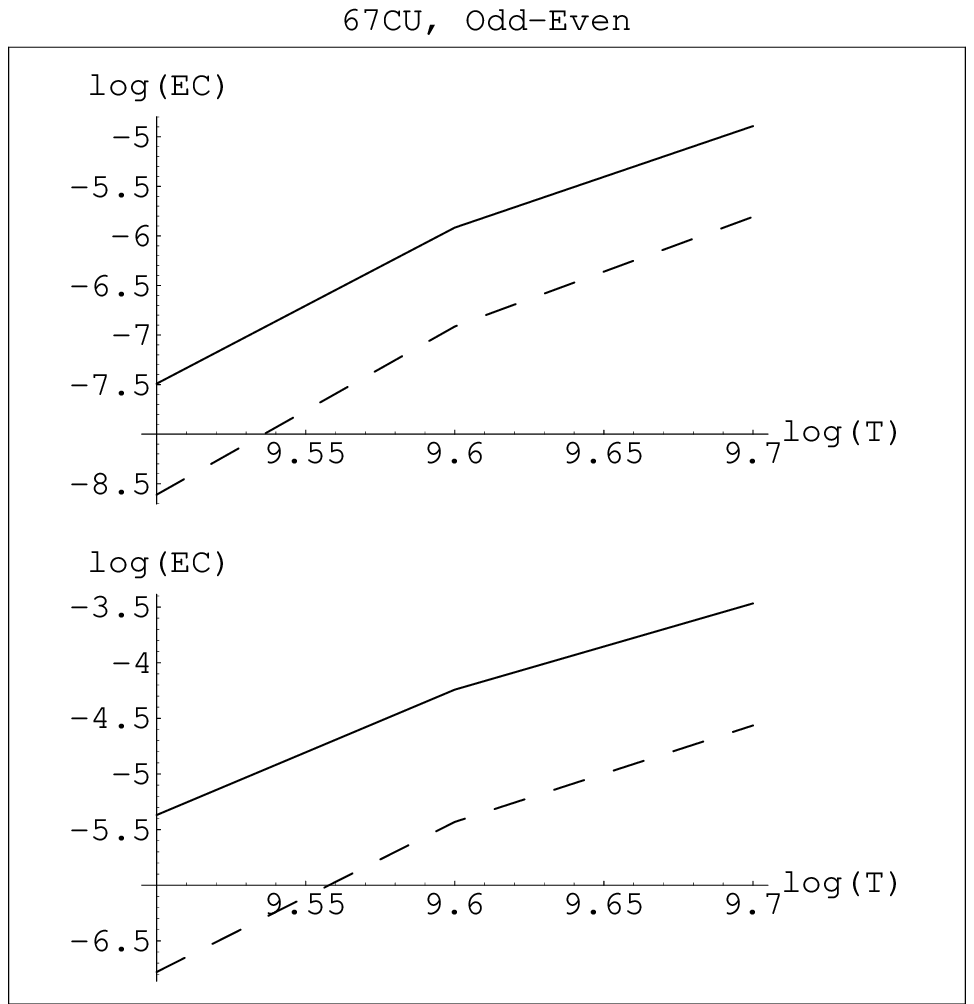}
\caption{ \footnotesize Same as Fig.~12 but for \ec by $^{67}$Cu.}
\end{figure}

Figures~12--14 represent the comparison of \ec rates of $^{62}$Co, $^{66}$Cu and $^{67}$Cu, respectively. ABKSV underestimated the total transition strength which they also pointed out in their paper \cite{Auf90}. Parent excited states till only 0.5061~MeV were considered by ABKSV for the case of $^{62}$Co (in comparison this work considers parent excited states up to 7.604~MeV). For $^{66}$Cu and $^{67}$Cu ABKSV considered only transitions from the ground-state. As can be easily seen their rates are suppressed since the \gt strengths from higher parent excited states are neglected. For $^{68}$Cu and $^{70}$Cu, ABKSV calculated the transition strengths from a method given in their Appendix A. They still considered transitions from the ground-state but calculated quite a  high value of the \gt strength (see Table~2 of \cite{Auf90}). Fig.~15 represents the situation for $^{68}$Cu. Electron capture rates for $^{70}$Cu were not stated as a function of temperature and density in their paper and hence no comparison could be made for this isotope of copper.
\begin{figure}
\epsfxsize=7.8cm
\epsffile{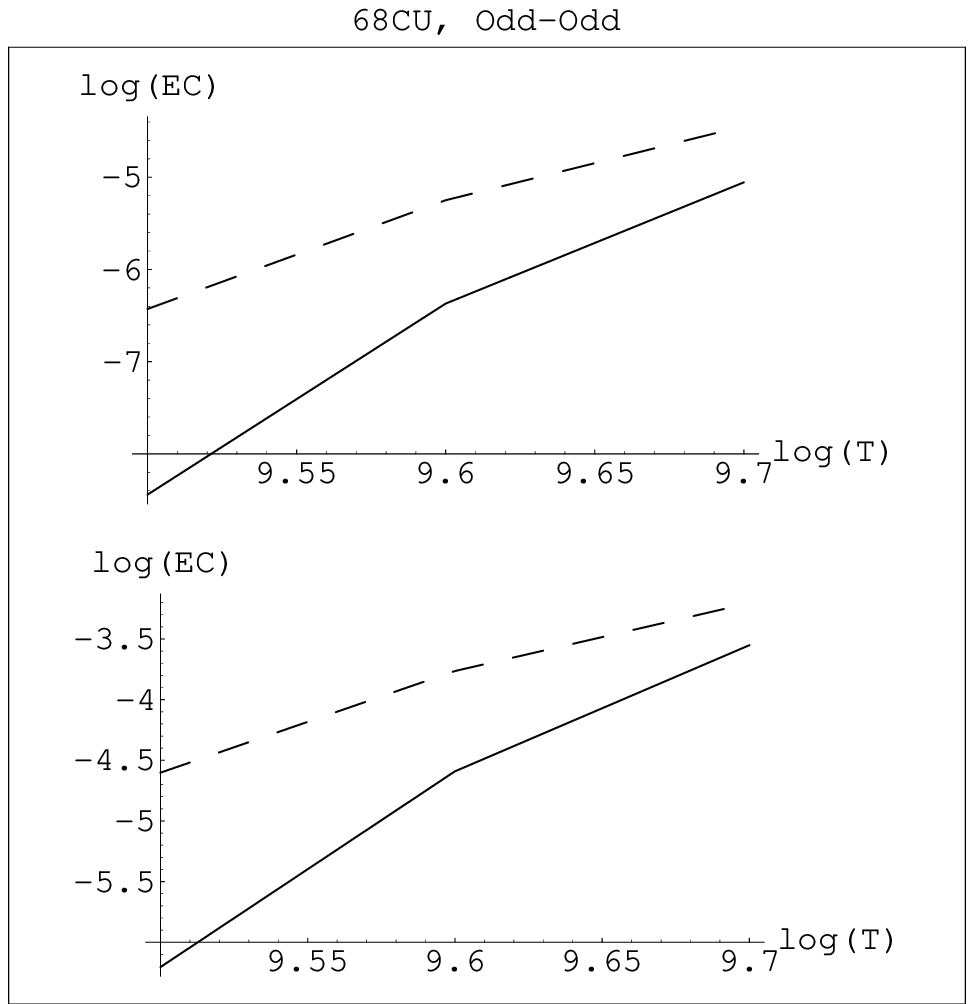}
\caption{ \footnotesize Same as Fig.~12 but for \ec by $^{68}$Cu.}
\end{figure}

A search for important weak interaction nuclei in presupernova evolution was made by AFWH. Tables~4--11 (in \cite{Auf94}) rank the most important \ec and beta-decay nuclei where rate calculations were done for specific values of temperature, density and $Y_{e}$. All of these 160 rates were re-calculated using the pn-QRPA theory and the comparison of the two calculations are presented in Tables~(2)--(9). It should be noted that AFWH calculated the \ec and \bt decay rates for $fp$-shell nuclei in the mass range 48 to 68.

\section{SUMMARY}
This paper, which is the third in a series starting from \cite{Nab99}, calculates microscopically for the first time the \wi rates in stellar environment for a total of 209 $fp$-shell nuclei in the mass range A = 61 to 80. This region of nuclei is a region of notorious complexity and the large number of excited states involved makes a full-blown shell model calculation a formidable task computationally. On the other hand various phenomenological parameterizations of the \gt strength of low-lying states were done previously for an insufficient number of heavier nuclei considering only a few parent excited states. The pn-QRPA theory is a good approach for tackling the problem of microscopic calculation of \wi rates for these nuclei.

The pn-QRPA theory earlier was quite successful in calculation of the terrestrial half-lives of beta-minus decay and beta-plus (\ec) decay and good comparison was achieved with the experimental data. The model works better with increasing neutron excess (see Table B and Fig.~6 of \cite{Sta90a}). This was a promising feature with respect to the prediction of unknown half-lives, implying that the predictions were made on the basis of a realistic physical model. Our calculations also include the \wi rates of many neutron-rich nuclei which play an important role in the core collapse of massive stars.

We plan in future to employ these microscopic \wi rates in simulations of stellar and galactic evolution processes and nucleosynthesis calculations.
\\
\\
\textit{ Acknowledgment}\\
One of the authors, J.-U. Nabi, will like to thank the Max-Planck-Society for the grant of their scholarship during the course of this work.   

\onecolumn
\noindent
\textbf{Table (1):} References of experimental data incorporated in this calculation (NP $\rightarrow$ Nuclear Physics, NDS $\rightarrow$ Nuclear Data Sheets).\\
\vspace*{0.1cm}\\
\begin{tabular}{|c|c|} \hline
Mass Number A & Reference \\ \hline
61      & NDS 67,271 (1992), NDS 38,463 (1983)\\
62      & NDS 60,337 (1990)\\
63      & NDS 64,815 (1991), NDS 28,559 (1979)\\
64      & NDS 78,395 (1996)\\
65      & NDS 69,209 (1993), NDS 47,135 (1986)\\
66      & NDS 83,789 (1998)\\
67      & NDS 64,875 (1991), NDS 39,741 (1983)\\
68      & NDS 76,343 (1995), NDS 55,1 (1988)\\
69      & NDS 58,1 (1989)\\
70      & NDS 68,117 (1993), NDS 51,95 (1987)\\
71      & NDS 68,579 (1993), NDS 53,1 (1988)\\
72      & NDS 73,215 (1994), NDS 56,1 (1989)\\
73      & NDS 69,857 (1993), NDS 51,161 (1987)\\
74      & NDS 74,529 (1995), NDS 51,225 (1987)\\
75      & NDS 60,735 (1990)\\
76      & NDS 74,63 (1995), NDS 42,233 (1984)\\
77      & NDS 81,417 (1997)\\
78      & NDS 63,1 (1991), NDS 33,189 (1981)\\ 
79      & NDS 70,437 (1993), NDS 37,393 (1982)\\
80      & NDS 66,623 (1992), NDS 36,127 (1982)\\\hline
\end{tabular}
\newpage
\begin{center}
\textbf{Comparison of Calculated Electron Capture  Rates with Those of Ref. [9]\\ (Table (2)--Table (5))}
\end{center}  
\textbf{\hspace*{3.5cm}Table (2)}\\
(Electron Capture Rates: $\rho = 5.86E+07, T9 = 3.40, Y_{e} = 0.47$)\\
\begin{tabular}{|c|c|c|} \hline
Nuclei & Rates (from [9]) & Rates (this work) \\ \hline
$^{57}$Co & 3.50E-03 &	4.56E-04 \\ 
$^{55}$Fe & 1.61E-03 & 	1.47E-03 \\
$^{55}$Co & 1.41E-01 &	3.99E-02 \\
$^{54}$Fe & 3.11E-04 &	3.07E-06 \\
$^{56}$Co & 7.40E-02 &	1.14E-02 \\
$^{53}$Mn & 2.48E-03 &	9.62E-04 \\
$^{58}$Ni & 6.36E-04 &	7.31E-05 \\
$^{59}$Ni & 4.37E-03 &	1.32E-03 \\
$^{61}$Cu & 3.93E-01 &	5.33E-02 \\
$^{57}$Ni & 1.94E-02 &	4.76E-02 \\
$^{58}$Co & 1.04E-02 &	2.23E-04 \\
$^{54}$Mn & 5.13E-03 &	4.34E-05 \\
$^{51}$Cr & 2.81E-03 &	7.19E-03 \\
$^{62}$Cu & 1.59E+00 &	1.68E-02 \\
$^{52}$Mn & 2.85E-02 &	2.52E-02 \\
$^{53}$Fe & 2.04E-02 &	5.08E-02 \\
$^{56}$Ni & 1.60E-02 & 	4.83E-03 \\
$^{59}$Cu & 1.01E+00 & 	1.42E-01 \\
$^{60}$Cu & 8.39E-01 &	1.29E-01 \\
$^{60}$Ni & 1.49E-06 &	1.70E-07 \\ \hline
\end{tabular}\\ 
\\ \\ 
\textbf{\hspace*{3.5cm}Table (3)}\\
(Electron Capture Rates: $\rho = 1.45E+08, T9 = 3.80, Y_{e} = 0.45$)\\
\begin{tabular}{|c|c|c|} \hline
Nuclei & Rates (from [9]) & Rates (this work) \\ \hline
$^{60}$Co & 1.27E-02 &	1.74E-05 \\ 
$^{59}$Co & 6.57E-04 & 	2.09E-04 \\
$^{54}$Mn & 1.57E-02 &	3.14E-04 \\
$^{64}$Cu & 3.69E-01 &	2.35E-02 \\
$^{61}$Ni & 1.20E-03 &	3.54E-04 \\
$^{63}$Cu & 1.86E-02 &	1.29E-02 \\
$^{58}$Co & 3.07E-02 &	1.55E-03 \\
$^{53}$Mn & 8.97E-03 &	5.64E-03 \\
$^{55}$Fe & 6.00E-03 &	6.42E-03 \\
$^{55}$Mn & 2.25E-05 &	9.62E-06 \\
$^{57}$Co & 1.29E-02 &	2.40E-03 \\
$^{57}$Fe & 1.84E-05 &	1.05E-05 \\
$^{62}$Cu & 4.60E+00 &	8.76E-02 \\
$^{51}$V  & 2.96E-05 &	1.27E-05 \\
$^{56}$Mn & 2.56E-04 &	2.42E-04 \\
$^{50}$V  & 2.45E-02 &	3.80E-03 \\
$^{56}$Fe & 1.31E-06 & 	7.14E-07 \\
$^{60}$Ni & 2.74E-05 & 	3.23E-06 \\
$^{51}$Cr & 9.33E-03 &	2.64E-02 \\
$^{53}$Cr & 2.46E-06 &	8.85E-06 \\ \hline
\end{tabular}\\
\textbf{\hspace*{3.5cm}Table (4)}\\
(Electron Capture Rates: $\rho = 1.06E+09, T9 = 4.93, Y_{e} = 0.43$)\\
\begin{tabular}{|c|c|c|} \hline
Nuclei & Rates (from [9]) & Rates (this work) \\ \hline
$^{60}$Co & 2.31E+00 &	1.36E-02 \\ 
$^{66}$Cu & 6.56E-01 & 	1.10E+00 \\
$^{62}$Co & 3.22E-02 &	9.34E-03 \\
$^{68}$Cu & 2.62E-02 &	1.56E-02 \\
$^{56}$Mn & 2.90E-02 &	2.77E-02 \\
$^{52}$V  & 1.81E-02 &	5.86E-04 \\
$^{59}$Co & 3.35E-01 &	3.99E-02 \\
$^{61}$Co & 2.23E-03 &	3.22E-03 \\
$^{67}$Cu & 2.38E-03 &	1.25E-02 \\
$^{65}$Cu & 5.91E-02 &	8.25E-02 \\
$^{51}$V  & 9.71E-03 &	5.78E-03 \\
$^{48}$Sc & 8.25E-02 &	4.68E-03 \\
$^{64}$Cu & 1.10E+01 &	2.51E+00 \\
$^{63}$Ni & 3.87E-03 &	2.47E-03 \\
$^{55}$Mn & 8.73E-03 &	5.89E-03 \\
$^{58}$Mn & 2.94E-03 &	3.23E-04 \\
$^{57}$Mn & 4.03E-04 & 	2.94E-04 \\
$^{59}$Fe & 1.83E-04 & 	5.70E-04 \\
$^{49}$Ti & 1.70E-02 &	7.17E-02 \\
$^{61}$Ni & 5.07E-01 &	3.46E-02 \\ \hline
\end{tabular}\\ 
\\ \\ \\ \\
\textbf{\hspace*{3.5cm}Table (5)}\\
(Electron Capture Rates: $\rho = 4.01E+10, T9 = 7.33, Y_{e} = 0.41$)\\
\begin{tabular}{|c|c|c|} \hline
Nuclei & Rates (from [9]) & Rates (this work) \\ \hline
$^{58}$Mn & 1.05E+03 &	7.45E+02 \\ 
$^{61}$Fe & 1.63E+02 & 	2.41E+02 \\
$^{49}$Sc & 5.23E+01 &	4.49E+02 \\
$^{63}$Co & 1.62E+02 &	3.69E+02 \\
$^{57}$Mn & 8.36E+02 &	3.41E+02 \\
$^{65}$Ni & 1.44E+02 &	4.23E+02 \\
$^{64}$Co & 2.40E+02 &	6.29E+02 \\
$^{55}$V  & 9.23E+01 &	5.52E+01 \\
$^{53}$V  & 1.73E+02 &	4.27E+02 \\
$^{59}$Mn & 1.41E+02 &	2.84E+02 \\
$^{60}$Mn & 2.55E+02 &	7.19E+02 \\
$^{57}$Cr & 6.09E+01 &	8.49E+01 \\
$^{59}$Fe & 7.20E+02 &	2.70E+02 \\
$^{62}$Co & 1.44E+03 &	5.23E+02 \\
$^{50}$Sc & 2.55E+01 &	7.39E+02 \\
$^{60}$Fe & 6.73E+01 &	3.02E+01 \\
$^{68}$Cu & 2.22E+02 & 	2.12E+03 \\
$^{54}$V  & 9.66E+01 & 	5.11E+02 \\
$^{56}$Cr & 3.33E+01 &	2.95E+01 \\
$^{61}$Co & 1.15E+03 &	3.40E+02 \\ \hline
\end{tabular}\\
\begin{center}
\textbf{Comparison of Calculated Beta Decay Rates with Those of Ref. [9]\\ (Table (6)--Table (9))}
\end{center} 
\textbf{\hspace*{3.5cm}Table (6)}\\
(Beta Decay Rates: $\rho = 5.86E+07, T9 = 3.40, Y_{e} = 0.47$)\\
\begin{tabular}{|c|c|c|} \hline
Nuclei & Rates (from [9]) & Rates (this work) \\ \hline
$^{57}$Fe & 1.10E-05 &	8.14E-08 \\ 
$^{54}$Mn & 8.81E-06 & 	7.17E-10 \\
$^{58}$Co & 5.67E-06 &	1.09E-10 \\
$^{53}$Cr & 5.93E-06 &	6.11E-07 \\
$^{60}$Co & 4.66E-03 &	1.97E-06 \\
$^{56}$Mn & 1.09E-02 &	1.86E-05 \\
$^{55}$Mn & 3.68E-07 &	2.01E-06 \\
$^{56}$Fe & 1.19E-10 &	1.72E-12 \\
$^{57}$Co & 1.34E-09 &	9.93E-12 \\
$^{59}$Co & 8.11E-08 &	8.65E-08 \\
$^{59}$Fe & 6.95E-03 &	3.00E-05 \\
$^{50}$V  & 1.89E-05 &	3.12E-09 \\
$^{56}$Co & 1.67E-08 &	0.00E+00 \\
$^{58}$Fe & 1.09E-07 &	1.35E-08 \\
$^{52}$V  & 1.60E-02 &	6.01E-04 \\
$^{55}$Fe & 1.30E-10 &	2.71E-12 \\
$^{61}$Co & 1.36E-03 & 	6.55E-04 \\
$^{63}$Ni & 3.16E-04 & 	4.56E-07 \\
$^{54}$Cr & 2.90E-07 &	3.36E-08 \\
$^{64}$Cu & 1.34E-04 &	2.76E-07 \\ \hline
\end{tabular}\\ 
\\ \\ 
\textbf{\hspace*{3.5cm}Table (7)}\\
(Beta Decay Rates: $\rho = 1.45E+08, T9 = 3.80, Y_{e} = 0.45$)\\
\begin{tabular}{|c|c|c|} \hline
Nuclei & Rates (from [9]) & Rates (this work) \\ \hline
$^{59}$Fe & 1.02E-02 &	7.53E-05 \\ 
$^{61}$Co & 1.90E-03 & 	3.46E-04 \\
$^{60}$Fe & 3.33E-03 &	1.93E-05 \\
$^{56}$Mn & 7.99E-03 &	1.25E-05 \\
$^{60}$Co & 4.54E-03 &	2.48E-06 \\
$^{52}$V  & 1.23E-02 &	3.24E-04 \\
$^{55}$Cr & 1.52E-03 &	3.40E-03 \\
$^{57}$Mn & 1.36E-03 &	7.99E-03 \\
$^{53}$V  & 5.60E-03 &	5.26E-03 \\
$^{62}$Co & 7.87E-02 &	6.09E-03 \\
$^{63}$Ni & 3.02E-04 &	3.65E-07 \\
$^{53}$Cr & 1.57E-05 &	7.99E-07 \\
$^{57}$Fe & 2.65E-05 &	1.26E-07 \\
$^{54}$Cr & 1.61E-06 &	8.36E-08 \\
$^{65}$Ni & 4.06E-03 &	5.15E-04 \\
$^{58}$Fe & 4.96E-07 &	3.13E-08 \\
$^{58}$Mn & 1.71E-01 & 	9.58E-02 \\
$^{51}$Ti & 1.07E-03 & 	1.45E-03 \\
$^{63}$Co & 2.05E-02 &	2.19E-01 \\
$^{61}$Fe & 8.64E-02 &	1.10E-02 \\ \hline
\end{tabular}\\
\textbf{\hspace*{3.5cm}Table (8)}\\
(Beta Decay Rates: $\rho = 1.06E+09, T9 = 4.93, Y_{e} = 0.43$)\\
\begin{tabular}{|c|c|c|} \hline
Nuclei & Rates (from [9]) & Rates (this work) \\ \hline
$^{61}$Fe & 1.25E-01 &	3.49E-03 \\ 
$^{57}$Cr & 3.31E-01 & 	4.22E-02 \\
$^{59}$Mn & 3.50E-01 &	6.32E-01 \\
$^{62}$Fe & 6.38E-02 &	1.56E-03 \\
$^{63}$Co & 2.01E-02 &	1.96E-02 \\
$^{67}$Ni & 1.71E-02 &	1.34E-03 \\
$^{58}$Cr & 6.02E-01 &	6.17E-03 \\
$^{58}$Mn & 1.61E-01 &	3.12E-02 \\
$^{50}$Sc & 1.10E-01 &	7.04E-04 \\
$^{65}$Co & 1.85E-01 &	1.47E+00 \\
$^{64}$Co & 2.29E-01 &	3.19E-01 \\
$^{54}$V  & 1.18E-01 &	9.85E-03 \\
$^{65}$Ni & 2.70E-03 &	2.36E-05 \\
$^{51}$Ti & 2.13E-03 &	6.60E-04 \\
$^{62}$Co & 7.16E-02 &	4.94E-04 \\
$^{55}$Cr & 3.31E-03 &	4.90E-04 \\
$^{68}$Ni & 2.84E-03 & 	1.84E-04 \\
$^{59}$Fe & 3.50E-03 & 	1.84E-04 \\
$^{60}$Fe & 3.63E-04 &	3.42E-05 \\
$^{63}$Fe & 7.18E-01 &	5.98E-02 \\ \hline
\end{tabular}\\ 
\\ \\ \\ \\
\textbf{\hspace*{3.5cm}Table (9)}\\
(Beta Decay Rates: $\rho = 4.01E+10, T9 = 7.33, Y_{e} = 0.41$)\\
\begin{tabular}{|c|c|c|} \hline
Nuclei & Rates (from [9]) & Rates (this work) \\ \hline
$^{52}$Sc & 1.09E-03 &	1.17E-06 \\ 
$^{62}$Mn & 7.78E-03 & 	4.63E-04 \\
$^{56}$V  & 2.02E-03 &	4.70E-06 \\
$^{53}$Sc & 1.56E-03 &	1.50E-04 \\
$^{51}$Sc & 6.04E-05 &	1.21E-06 \\
$^{57}$V  & 1.84E-03 &	6.33E-05 \\
$^{60}$Mn & 4.40E-04 &	1.74E-06 \\
$^{51}$Ca & 1.47E-03 &	1.26E-05 \\
$^{55}$Ti & 9.70E-04 &	7.64E-05 \\
$^{68}$Co & 6.98E-04 &	1.20E-03 \\
$^{61}$Mn & 1.86E-04 &	5.11E-05 \\
$^{49}$Ca & 8.80E-06 &	3.23E-07 \\
$^{50}$Ca & 4.46E-05 &	1.27E-06 \\
$^{55}$V  & 5.75E-05 &	1.20E-05 \\
$^{59}$Cr & 3.50E-04 &	3.36E-05 \\
$^{63}$Fe & 1.17E-04 &	6.46E-07 \\
$^{53}$Ti & 1.53E-05 & 	2.46E-06 \\
$^{54}$Ti & 4.64E-05 & 	3.30E-06 \\
$^{50}$Sc & 1.61E-05 &	1.16E-08 \\
$^{66}$Co & 3.43E-05 &	5.14E-05 \\ \hline
\end{tabular}
\normalsize
\begin{center}
\textbf{\Large EXPLANATION OF TABLE}
\end{center}
\vspace{0.7in}
\textbf{\large TABLE A. $\mathbf{fp}$-Shell Nuclei Weak Rates in Stellar Matter}
\vspace{0.4in}
\noindent
\newline
The calculated weak interaction rates [Eqs. (15) and (18)-(20)] are all tabulated in log$_{10} \lambda$. The probabilities of $\beta$-delayed proton (neutron) emission [Eq.~(21)] are also tabulated in logarithmic scale. These probabilities are calculated only to one significant figure and are given up to three places of decimal only for designing purposes. All rates listed for a particular direction concern the parent nucleus except for the last two columns which concern the daughter nucleus. For each daughter nucleus, either the proton energy rate and probability of $\beta$-delayed proton emission is stated (if $S_{p} < S_{n}$) or the neutron energy rate and probability of $\beta$-delayed neutron emission is stated (if $S_{n} < S_{p}$). In the table -100 means that the rate (or the probability) is smaller than $10^{-100}$. \\
\begin {tabbing}
\textsf{Q} \hspace{0.7in}        \= Mass of parent minus mass of daughter nucleus\\
S$_{\mathsf{nP}}$                       \> Separation energy of neutron for parent nucleus\\
S$_{\mathsf{pP}}$                       \> Separation energy of proton for parent nucleus\\
S$_{\mathsf{nD}}$                       \> Separation energy of neutron for daughter nucleus\\
S$_{\mathsf{pD}}$                     \> Separation energy of neutron for daughter nucleus\\
\textsf{ADen}                    \> log $(\rho Y_{e})$ (g.cm$^{-3}$), where $\rho$ is the density of the baryon, and $Y_{e}$ is the ratio of the\\
                        \> electron number to the baryon number\\
\textsf{T9}                      \> Temperature in units of $10^{9}$ K\\
\textsf{EFermi}                  \> Total Fermi energy of electron and positron, including the rest mass (MeV)\\
\textsf{E+Cap}                   \> Positron capture rate (s$^{-1}$)\\
\textsf{E-Dec}                   \> Electron decay rate (s$^{-1}$)\\ 
\textsf{ANuEn}                   \> Anti-Neutrino energy loss rate (MeV.s$^{-1}$)\\
\textsf{GamEn}                 \> Gamma ray heating rate (MeV.s$^{-1}$), tabulated separately for $\beta^{+}$ direction \\
                        \> and $\beta^{-}$ direction \\
\textsf{E-Cap}                   \> Electron capture rate (s$^{-1}$)\\
\textsf{E+Dec}                   \> Positron decay rate (s$^{-1}$)\\
\textsf{NuEn}                 \> Neutrino energy loss rate (MeV.s$^{-1}$)\\     
\textsf{ProEn}                  \> Energy rate of $\beta$-delayed proton (MeV.s$^{-1}$)\\
\textsf{NeuEn}                  \> Energy rate of $\beta$-delayed neutron (MeV.s$^{-1}$)\\
\textsf{PPEm}                  \> Probability of $\beta$-delayed proton emission\\
\textsf{PNEm}                  \> Probability of $\beta$-delayed neutron emission\\             
\end{tabbing}
\clearpage
\sffamily \normalsize
\noindent
\begin{tabular}{rr|rrrrrr}
\multicolumn{7}{c}{\normalsize{\hspace*{3.5cm}$^{  61}_{23}$ V  $_{ 38} \rightarrow \hspace*{0.2cm}^{  61}_{24}$ CR $_{ 37}$}}\\
\hline
Aden & T9& E+Cap &   E-Dec &  ANuEn &  GamEn &  NeuEn &  PNEm \\
\hline
     3.0&  1.0 &  -5.026 &   1.724 &   2.506 &   1.463 &   0.610 & -1.000 \\
     3.0&  3.0 &  -1.971 &   1.648 &   2.421 &   1.362 &   0.850 & -0.900 \\
     3.0& 30.0 &   2.992 &   2.161 &   4.193 &   2.536 &   4.000 & -0.040 \\
     7.0&  1.0 & -10.846 &   1.720 &   2.499 &   1.458 &   0.565 & -1.000 \\
     7.0&  3.0 &  -3.675 &   1.643 &   2.414 &   1.357 &   0.820 & -1.000 \\
     7.0& 30.0 &   2.989 &   2.161 &   4.190 &   2.535 &   3.997 & -0.040 \\
    11.0&  1.0 &    -100 & -58.842 & -59.434 & -60.415 & -84.355 &-20.000 \\
    11.0&  3.0 & -42.152 & -20.376 & -20.503 & -21.224 & -28.351 & -7.000 \\
    11.0& 30.0 &  -0.858 &  -0.542 &   0.515 &  -0.191 &   0.219 & -0.200 \\
\hline
\end{tabular}\\
\vspace{0.20cm}\\
\begin{tabular}{rr|rrrrrr}
\multicolumn{7}{c}{\normalsize{\hspace*{3.5cm}$^{  61}_{23}$ V  $_{ 38} \rightarrow \hspace*{0.2cm}^{  61}_{22}$ TI $_{ 39}$}}\\
\hline
Aden & T9& E-Cap &   E+Dec &  NuEn &  GamEn &  NeuEn &  PNEm \\
\hline
     3.0&  1.0 & -74.359 &    -100 & -74.941 & -75.584 & -90.704 &-20.000 \\
     3.0&  3.0 & -25.580 &    -100 & -25.673 & -26.488 & -30.370 & -4.000 \\
     3.0& 30.0 &   0.556 &    -100 &   1.506 &  -1.796 &   1.337 & -0.004 \\
     7.0&  1.0 & -68.539 &    -100 & -69.121 & -69.764 & -84.884 &-20.000 \\
     7.0&  3.0 & -23.865 &    -100 & -23.958 & -24.772 & -28.654 & -4.000 \\
     7.0& 30.0 &   0.559 &    -100 &   1.510 &  -1.793 &   1.340 & -0.004 \\
    11.0&  1.0 &   3.142 &    -100 &   3.637 &   1.088 &   3.570 & -0.010 \\
    11.0&  3.0 &   3.124 &    -100 &   3.653 &   1.286 &   3.546 & -0.020 \\
    11.0& 30.0 &   4.299 &    -100 &   5.302 &   1.716 &   5.100 & -0.002 \\
\hline
\end{tabular}\\
\vspace{0.20cm}\\
\noindent
\begin{tabular}{rr|rrrrrr}
\multicolumn{7}{c}{\normalsize{\hspace*{3.5cm}$^{  61}_{24}$ CR $_{ 37} \rightarrow \hspace*{0.2cm}^{  61}_{25}$ MN $_{ 36}$}}\\
\hline
Aden & T9& E+Cap &   E-Dec &  ANuEn &  GamEn &  NeuEn &  PNEm \\
\hline
     3.0&  1.0 &  -5.679 &   0.416 &   1.027 &   0.551 &  -3.921 & -3.000 \\
     3.0&  3.0 &  -2.594 &   0.430 &   1.043 &   0.582 &  -3.276 & -3.000 \\
     3.0& 30.0 &   2.268 &   0.915 &   3.392 &   2.297 &   3.054 & -0.100 \\
     7.0&  1.0 & -11.499 &   0.401 &   1.006 &   0.515 &  -5.435 & -4.000 \\
     7.0&  3.0 &  -4.299 &   0.417 &   1.026 &   0.553 &  -3.772 & -4.000 \\
     7.0& 30.0 &   2.265 &   0.914 &   3.389 &   2.294 &   3.050 & -0.100 \\
    11.0&  1.0 &    -100 & -76.406 & -77.000 & -79.130 &    -100 &-40.000 \\
    11.0&  3.0 & -42.777 & -26.525 & -26.657 & -28.212 & -40.006 &-10.000 \\
    11.0& 30.0 &  -1.588 &  -1.989 &  -0.428 &  -1.249 &  -0.808 & -0.200 \\
\hline
\end{tabular}\\
\vspace{0.20cm}\\
\begin{tabular}{rr|rrrrrr}
\multicolumn{7}{c}{\normalsize{\hspace*{3.5cm}$^{  61}_{24}$ CR $_{ 37} \rightarrow \hspace*{0.2cm}^{  61}_{23}$ V  $_{ 38}$}}\\
\hline
Aden & T9& E-Cap &   E+Dec &  NuEn &  GamEn &  NeuEn &  PNEm \\
\hline
     3.0&  1.0 & -68.728 &    -100 & -69.310 & -70.178 &    -100 &-30.000 \\
     3.0&  3.0 & -23.916 &    -100 & -24.009 & -24.755 & -34.361 &-10.000 \\
     3.0& 30.0 &  -0.335 &    -100 &   0.629 &   0.073 &  -0.433 & -0.500 \\
     7.0&  1.0 & -62.908 &    -100 & -63.490 & -64.358 & -95.404 &-30.000 \\
     7.0&  3.0 & -22.200 &    -100 & -22.294 & -23.040 & -32.645 &-10.000 \\
     7.0& 30.0 &  -0.332 &    -100 &   0.633 &   0.077 &  -0.430 & -0.500 \\
    11.0&  1.0 &   2.532 &    -100 &   3.310 &   3.022 &   1.001 & -2.000 \\
    11.0&  3.0 &   2.570 &    -100 &   3.364 &   3.040 &   1.093 & -2.000 \\
    11.0& 30.0 &   3.274 &    -100 &   4.335 &   3.645 &   3.310 & -0.400 \\
\hline
\end{tabular}\\
\vspace{0.20cm}\\
\noindent
\begin{tabular}{rr|rrrrrr}
\multicolumn{7}{c}{\normalsize{\hspace*{3.5cm}$^{  61}_{25}$ MN $_{ 36} \rightarrow \hspace*{0.2cm}^{  61}_{26}$ FE $_{ 35}$}}\\
\hline
Aden & T9& E+Cap &   E-Dec &  ANuEn &  GamEn &  NeuEn &  PNEm \\
\hline
     3.0&  1.0 &  -5.026 &   1.306 &   1.868 &   0.933 &  -6.666 & -7.000 \\
     3.0&  3.0 &  -2.078 &   1.206 &   1.767 &   0.839 &  -4.508 & -6.000 \\
     3.0& 30.0 &   2.597 &   1.813 &   3.848 &   3.212 &   2.748 & -0.600 \\
     7.0&  1.0 & -10.846 &   1.292 &   1.843 &   0.917 &  -8.645 & -9.000 \\
     7.0&  3.0 &  -3.783 &   1.193 &   1.746 &   0.825 &  -5.593 & -6.000 \\
     7.0& 30.0 &   2.594 &   1.813 &   3.845 &   3.209 &   2.744 & -0.600 \\
    11.0&  1.0 &    -100 & -85.593 & -86.188 & -86.911 &    -100 &-30.000 \\
    11.0&  3.0 & -42.260 & -29.242 & -29.382 & -29.799 & -39.493 &-10.000 \\
    11.0& 30.0 &  -1.248 &  -1.267 &   0.050 &  -0.372 &  -1.081 & -0.700 \\
\hline
\end{tabular}\\
\vspace{0.20cm}\\
\begin{tabular}{rr|rrrrrr}
\multicolumn{7}{c}{\normalsize{\hspace*{3.5cm}$^{  61}_{25}$ MN $_{ 36} \rightarrow \hspace*{0.2cm}^{  61}_{24}$ CR $_{ 37}$}}\\
\hline
Aden & T9& E-Cap &   E+Dec &  NuEn &  GamEn &  NeuEn &  PNEm \\
\hline
     3.0&  1.0 & -49.626 &    -100 & -50.206 & -51.722 & -75.192 &-20.000 \\
     3.0&  3.0 & -16.551 &    -100 & -16.640 & -17.702 & -24.435 & -7.000 \\
     3.0& 30.0 &   1.995 &    -100 &   2.997 &   2.428 &   1.291 & -0.700 \\
     7.0&  1.0 & -43.806 &    -100 & -44.386 & -45.902 & -69.372 &-20.000 \\
     7.0&  3.0 & -14.836 &    -100 & -14.924 & -15.987 & -22.719 & -7.000 \\
     7.0& 30.0 &   1.998 &    -100 &   3.001 &   2.431 &   1.295 & -0.700 \\
    11.0&  1.0 &   4.538 &    -100 &   5.586 &   4.061 &   2.607 & -2.000 \\
    11.0&  3.0 &   4.436 &    -100 &   5.484 &   3.986 &   2.623 & -2.000 \\
    11.0& 30.0 &   5.225 &    -100 &   6.427 &   5.655 &   4.661 & -0.600 \\
\hline
\end{tabular}\\
\vspace{0.20cm}\\
\noindent
\begin{tabular}{rr|rrrrrr}
\multicolumn{7}{c}{\normalsize{\hspace*{3.5cm}$^{  61}_{26}$ FE $_{ 35} \rightarrow \hspace*{0.2cm}^{  61}_{27}$ CO $_{ 34}$}}\\
\hline
Aden& T9&  E+Cap &   E-Dec &  ANuEn &  GamEn &  ProEn &  PPEm \\
\hline
     3.0&  1.0 &  -7.680 &  -2.435 &  -2.163 &  -2.565 & -35.168 &-30.000 \\
     3.0&  3.0 &  -4.383 &  -1.921 &  -1.554 &  -2.448 & -12.712 &-10.000 \\
     3.0& 30.0 &   1.843 &   0.057 &   3.007 &   2.513 &   1.899 & -0.700 \\
     7.0&  1.0 & -13.499 &  -2.522 &  -2.289 &  -2.686 & -39.510 &-40.000 \\
     7.0&  3.0 &  -6.088 &  -1.963 &  -1.618 &  -2.533 & -14.424 &-10.000 \\
     7.0& 30.0 &   1.840 &   0.056 &   3.004 &   2.509 &   1.895 & -0.700 \\
    11.0&  1.0 &    -100 &    -100 &    -100 &    -100 &    -100 &-100\\
    11.0&  3.0 & -44.565 & -35.348 & -35.498 & -37.246 & -52.901 &-20.000 \\
    11.0& 30.0 &  -2.007 &  -3.222 &  -0.846 &  -1.318 &  -1.966 & -0.700 \\
\hline
\end{tabular}\\
\vspace{0.20cm}\\
\begin{tabular}{rr|rrrrrr}
\multicolumn{7}{c}{\normalsize{\hspace*{3.5cm}$^{  61}_{26}$ FE $_{ 35} \rightarrow \hspace*{0.2cm}^{  61}_{25}$ MN $_{ 36}$}}\\
\hline
Aden & T9& E-Cap &   E+Dec &  NuEn &  GamEn &  NeuEn &  PNEm \\
\hline
     3.0&  1.0 & -42.274 &    -100 & -42.852 & -43.545 & -82.684 &-40.000 \\
     3.0&  3.0 & -14.532 &    -100 & -14.618 & -14.818 & -28.190 &-10.000 \\
     3.0& 30.0 &   1.722 &    -100 &   2.725 &   2.391 &   0.315 & -2.000 \\
     7.0&  1.0 & -36.455 &    -100 & -37.032 & -37.725 & -76.864 &-40.000 \\
     7.0&  3.0 & -12.817 &    -100 & -12.903 & -13.103 & -26.474 &-10.000 \\
     7.0& 30.0 &   1.726 &    -100 &   2.729 &   2.394 &   0.318 & -2.000 \\
    11.0&  1.0 &   3.660 &    -100 &   4.735 &   3.790 &   1.726 & -2.000 \\
    11.0&  3.0 &   4.083 &    -100 &   5.178 &   4.045 &   1.830 & -2.000 \\
    11.0& 30.0 &   4.940 &    -100 &   6.145 &   5.603 &   3.887 & -2.000 \\
\hline
\end{tabular}\\
\vspace{0.20cm}\\
\noindent
\begin{tabular}{rr|rrrrrr}
\multicolumn{7}{c}{\normalsize{\hspace*{3.5cm}$^{  61}_{27}$ CO $_{ 34} \rightarrow \hspace*{0.2cm}^{  61}_{28}$ NI $_{ 33}$}}\\
\hline
Aden & T9& E+Cap &   E-Dec &  ANuEn &  GamEn &  NeuEn &  PNEm \\
\hline
     3.0&  1.0 &  -7.163 &  -3.231 &  -3.393 &  -3.768 & -42.025 &-40.000 \\
     3.0&  3.0 &  -3.713 &  -2.813 &  -2.720 &  -3.317 & -14.232 &-10.000 \\
     3.0& 30.0 &   2.264 &  -0.535 &   3.348 &   2.532 &   2.781 & -0.200 \\
     7.0&  1.0 & -12.983 &  -4.004 &  -4.405 &  -4.586 & -47.180 &-40.000 \\
     7.0&  3.0 &  -5.421 &  -3.060 &  -3.109 &  -3.887 & -15.929 &-10.000 \\
     7.0& 30.0 &   2.261 &  -0.535 &   3.344 &   2.529 &   2.777 & -0.200 \\
    11.0&  1.0 &    -100 &    -100 &    -100 &    -100 &    -100 &-100\\
    11.0&  3.0 & -43.898 & -39.809 & -39.983 & -40.571 & -54.351 &-10.000 \\
    11.0& 30.0 &  -1.594 &  -3.788 &  -0.514 &  -1.317 &  -1.083 & -0.200 \\
\hline
\end{tabular}\\
\vspace{0.20cm}\\
\begin{tabular}{rr|rrrrrr}
\multicolumn{7}{c}{\normalsize{\hspace*{3.5cm}$^{  61}_{27}$ CO $_{ 34} \rightarrow \hspace*{0.2cm}^{  61}_{26}$ FE $_{ 35}$}}\\
\hline
Aden & T9& E-Cap &   E+Dec &  NuEn &  GamEn &  NeuEn &  PNEm \\
\hline
     3.0&  1.0 & -26.964 & -34.510 & -27.518 & -28.303 & -59.221 &-30.000 \\
     3.0&  3.0 & -10.180 & -15.616 & -10.220 & -10.759 & -19.683 & -9.000 \\
     3.0& 30.0 &   2.035 &  -7.676 &   3.055 &   2.155 &   2.437 & -0.100 \\
     7.0&  1.0 & -21.146 & -34.510 & -21.699 & -22.484 & -53.401 &-30.000 \\
     7.0&  3.0 &  -8.467 & -15.593 &  -8.506 &  -9.046 & -17.967 & -9.000 \\
     7.0& 30.0 &   2.038 &  -7.675 &   3.058 &   2.159 &   2.441 & -0.100 \\
    11.0&  1.0 &   4.400 & -34.510 &   5.502 &   4.942 &   2.712 & -2.000 \\
    11.0&  3.0 &   4.390 & -15.593 &   5.493 &   4.940 &   2.736 & -2.000 \\
    11.0& 30.0 &   5.189 &  -7.435 &   6.410 &   5.145 &   5.666 & -0.090 \\
\hline
\end{tabular}\\
\vspace{0.20cm}\\
\noindent
\begin{tabular}{rr|rrrrrr}
\multicolumn{7}{c}{\normalsize{\hspace*{3.5cm}$^{  61}_{28}$ NI $_{ 33} \rightarrow \hspace*{0.2cm}^{  61}_{29}$ CU $_{ 32}$}}\\
\hline
Aden& T9&  E+Cap &   E-Dec &  ANuEn &  GamEn &  ProEn &  PPEm \\
\hline
     3.0&  1.0 & -17.753 & -19.699 & -17.726 & -19.558 & -46.147 &-30.000 \\
     3.0&  3.0 &  -7.692 & -10.108 &  -7.510 &  -8.761 & -16.154 & -8.000 \\
     3.0& 30.0 &   1.196 &  -3.690 &   2.230 &   0.906 &   2.013 & -0.060 \\
     7.0&  1.0 & -23.573 & -21.704 & -22.323 & -23.189 & -51.421 &-30.000 \\
     7.0&  3.0 &  -9.407 & -10.598 &  -9.217 & -10.429 & -17.869 & -8.000 \\
     7.0& 30.0 &   1.193 &  -3.691 &   2.227 &   0.903 &   2.010 & -0.060 \\
    11.0&  1.0 &    -100 &    -100 &    -100 &    -100 &    -100 &-100\\
    11.0&  3.0 & -47.884 & -48.787 & -47.688 & -48.837 & -56.346 & -8.000 \\
    11.0& 30.0 &  -2.667 &  -7.077 &  -1.634 &  -2.953 &  -1.852 & -0.060 \\
\hline
\end{tabular}\\
\vspace{0.20cm}\\
\begin{tabular}{rr|rrrrrr}
\multicolumn{7}{c}{\normalsize{\hspace*{3.5cm}$^{  61}_{28}$ NI $_{ 33} \rightarrow \hspace*{0.2cm}^{  61}_{27}$ CO $_{ 34}$}}\\
\hline
Aden& T9&  E-Cap &   E+Dec &  NuEn &  GamEn &  ProEn &  PPEm \\
\hline
     3.0&  1.0 & -14.635 & -20.412 & -15.172 & -16.092 & -61.935 &-50.000 \\
     3.0&  3.0 &  -6.943 & -10.666 &  -6.936 &  -7.833 & -20.785 &-10.000 \\
     3.0& 30.0 &   1.749 &  -4.127 &   2.765 &   2.078 &   2.112 & -0.200 \\
     7.0&  1.0 &  -8.943 & -20.412 &  -9.445 & -10.344 & -56.115 &-50.000 \\
     7.0&  3.0 &  -5.254 & -10.645 &  -5.241 &  -6.135 & -19.069 &-10.000 \\
     7.0& 30.0 &   1.753 &  -4.126 &   2.768 &   2.081 &   2.116 & -0.200 \\
    11.0&  1.0 &   4.362 & -20.412 &   5.469 &   5.144 &   2.011 & -3.000 \\
    11.0&  3.0 &   4.256 & -10.645 &   5.365 &   5.045 &   1.947 & -3.000 \\
    11.0& 30.0 &   4.927 &  -3.945 &   6.141 &   5.173 &   5.348 & -0.100 \\
\hline
\end{tabular}\\
\vspace{0.20cm}\\
\noindent
\begin{tabular}{rr|rrrrrr}
\multicolumn{7}{c}{\normalsize{\hspace*{3.5cm}$^{  61}_{29}$ CU $_{ 32} \rightarrow \hspace*{0.2cm}^{  61}_{30}$ ZN $_{ 31}$}}\\
\hline
Aden& T9&  E+Cap &   E-Dec &  ANuEn &  GamEn &  ProEn &  PPEm \\
\hline
     3.0&  1.0 & -33.768 &    -100 & -33.716 & -34.667 & -64.709 &-30.000 \\
     3.0&  3.0 & -11.668 &    -100 & -11.506 & -12.047 & -20.989 & -9.000 \\
     3.0& 30.0 &   1.405 &    -100 &   2.429 &   1.443 &   1.601 & -0.200 \\
     7.0&  1.0 & -39.588 &    -100 & -39.536 & -40.487 & -70.529 &-30.000 \\
     7.0&  3.0 & -13.383 &    -100 & -13.221 & -13.762 & -22.705 & -9.000 \\
     7.0& 30.0 &   1.402 &    -100 &   2.425 &   1.440 &   1.598 & -0.200 \\
    11.0&  1.0 &    -100 &    -100 &    -100 &    -100 &    -100 &-100\\
    11.0&  3.0 & -51.860 &    -100 & -51.699 & -52.240 & -61.182 & -9.000 \\
    11.0& 30.0 &  -2.459 &    -100 &  -1.437 &  -2.421 &  -2.265 & -0.200 \\
\hline
\end{tabular}\\
\vspace{0.20cm}\\
\begin{tabular}{rr|rrrrrr}
\multicolumn{7}{c}{\normalsize{\hspace*{3.5cm}$^{  61}_{29}$ CU $_{ 32} \rightarrow \hspace*{0.2cm}^{  61}_{28}$ NI $_{ 33}$}}\\
\hline
Aden & T9& E-Cap &   E+Dec &  NuEn &  GamEn &  NeuEn &  PNEm \\
\hline
     3.0&  1.0 &  -6.198 &  -4.448 &  -4.606 &  -5.631 & -39.703 &-30.000 \\
     3.0&  3.0 &  -3.327 &  -3.286 &  -2.717 &  -3.277 & -13.617 &-10.000 \\
     3.0& 30.0 &   2.286 &  -1.417 &   3.380 &   2.822 &   2.061 & -0.600 \\
     7.0&  1.0 &  -2.382 &  -4.448 &  -2.029 &  -2.657 & -33.883 &-30.000 \\
     7.0&  3.0 &  -1.859 &  -3.282 &  -1.400 &  -1.888 & -11.902 &-10.000 \\
     7.0& 30.0 &   2.289 &  -1.416 &   3.384 &   2.826 &   2.064 & -0.600 \\
    11.0&  1.0 &   4.656 &  -4.448 &   5.849 &   5.320 &   4.164 & -0.800 \\
    11.0&  3.0 &   4.702 &  -3.281 &   5.908 &   5.352 &   4.185 & -0.800 \\
    11.0& 30.0 &   5.154 &  -1.265 &   6.466 &   5.656 &   5.185 & -0.400 \\
\hline
\end{tabular}\\
\vspace{0.20cm}\\
\noindent
\begin{tabular}{rr|rrrrrr}
\multicolumn{7}{c}{\normalsize{\hspace*{3.5cm}$^{  61}_{30}$ ZN $_{ 31} \rightarrow \hspace*{0.2cm}^{  61}_{31}$ GA $_{ 30}$}}\\
\hline
Aden& T9&  E+Cap &   E-Dec &  ANuEn &  GamEn &  ProEn &  PPEm \\
\hline
     3.0&  1.0 & -50.439 &    -100 & -50.388 & -52.743 & -59.343 & -9.000 \\
     3.0&  3.0 & -17.565 &    -100 & -17.410 & -18.208 & -20.282 & -2.000 \\
     3.0& 30.0 &   0.617 &    -100 &   1.611 &  -1.093 &   1.479 & -0.020 \\
     7.0&  1.0 & -56.259 &    -100 & -56.208 & -58.563 & -65.163 & -9.000 \\
     7.0&  3.0 & -19.281 &    -100 & -19.125 & -19.923 & -21.998 & -2.000 \\
     7.0& 30.0 &   0.613 &    -100 &   1.607 &  -1.096 &   1.476 & -0.020 \\
    11.0&  1.0 &    -100 &    -100 &    -100 &    -100 &    -100 &-100\\
    11.0&  3.0 & -57.758 &    -100 & -57.603 & -58.401 & -60.475 & -2.000 \\
    11.0& 30.0 &  -3.249 &    -100 &  -2.255 &  -4.958 &  -2.387 & -0.020 \\
\hline
\end{tabular}\\
\vspace{0.20cm}\\
\begin{tabular}{rr|rrrrrr}
\multicolumn{7}{c}{\normalsize{\hspace*{3.5cm}$^{  61}_{30}$ ZN $_{ 31} \rightarrow \hspace*{0.2cm}^{  61}_{29}$ CU $_{ 32}$}}\\
\hline
Aden& T9&  E-Cap &   E+Dec &  NuEn &  GamEn &  ProEn &  PPEm \\
\hline
     3.0&  1.0 &  -5.325 &  -1.960 &  -1.601 &  -2.265 & -10.051 & -7.000 \\
     3.0&  3.0 &  -2.701 &  -1.491 &  -1.081 &  -1.616 &  -5.191 & -3.000 \\
     3.0& 30.0 &   2.569 &  -0.527 &   3.662 &   2.291 &   3.250 & -0.060 \\
     7.0&  1.0 &  -1.563 &  -1.960 &  -0.910 &  -1.116 &  -4.673 & -3.000 \\
     7.0&  3.0 &  -1.260 &  -1.490 &  -0.488 &  -0.878 &  -3.514 & -2.000 \\
     7.0& 30.0 &   2.573 &  -0.526 &   3.665 &   2.295 &   3.254 & -0.060 \\
    11.0&  1.0 &   5.059 &  -1.960 &   6.286 &   4.640 &   5.503 & -0.040 \\
    11.0&  3.0 &   5.100 &  -1.490 &   6.330 &   4.669 &   5.559 & -0.050 \\
    11.0& 30.0 &   5.443 &  -0.397 &   6.753 &   4.915 &   6.207 & -0.030 \\
\hline
\end{tabular}\\
\vspace{0.20cm}\\
\noindent
\begin{tabular}{rr|rrrrrr}
\multicolumn{7}{c}{\normalsize{\hspace*{3.5cm}$^{  61}_{31}$ GA $_{ 30} \rightarrow \hspace*{0.2cm}^{  61}_{32}$ GE $_{ 29}$}}\\
\hline
Aden& T9&  E+Cap &   E-Dec &  ANuEn &  GamEn &  ProEn &  PPEm \\
\hline
     3.0&  1.0 & -72.813 &    -100 & -72.762 & -75.039 & -86.507 &-10.000 \\
     3.0&  3.0 & -24.459 &    -100 & -24.307 & -25.591 & -29.386 & -5.000 \\
     3.0& 30.0 &  -0.154 &    -100 &   0.820 &  -1.409 &   0.365 & -0.100 \\
     7.0&  1.0 & -78.633 &    -100 & -78.582 & -80.859 & -92.327 &-10.000 \\
     7.0&  3.0 & -26.174 &    -100 & -26.022 & -27.307 & -31.102 & -5.000 \\
     7.0& 30.0 &  -0.158 &    -100 &   0.816 &  -1.412 &   0.362 & -0.100 \\
    11.0&  1.0 &    -100 &    -100 &    -100 &    -100 &    -100 &-100\\
    11.0&  3.0 & -64.652 &    -100 & -64.500 & -65.784 & -69.579 & -5.000 \\
    11.0& 30.0 &  -4.021 &    -100 &  -3.047 &  -5.275 &  -3.502 & -0.100 \\
\hline
\end{tabular}\\
\vspace{0.20cm}\\
\begin{tabular}{rr|rrrrrr}
\multicolumn{7}{c}{\normalsize{\hspace*{3.5cm}$^{  61}_{31}$ GA $_{ 30} \rightarrow \hspace*{0.2cm}^{  61}_{30}$ ZN $_{ 31}$}}\\
\hline
Aden& T9&  E-Cap &   E+Dec &  NuEn &  GamEn &  ProEn &  PPEm \\
\hline
     3.0&  1.0 &  -3.950 &  -0.062 &   0.514 &  -0.134 &  -3.044 & -3.000 \\
     3.0&  3.0 &  -1.569 &   0.046 &   0.642 &   0.019 &  -1.859 & -2.000 \\
     3.0& 30.0 &   2.667 &   0.088 &   3.808 &   2.626 &   2.776 & -0.100 \\
     7.0&  1.0 &  -0.210 &  -0.062 &   0.780 &   0.250 &  -0.545 & -0.700 \\
     7.0&  3.0 &  -0.128 &   0.047 &   0.896 &   0.346 &  -0.417 & -0.700 \\
     7.0& 30.0 &   2.670 &   0.088 &   3.812 &   2.629 &   2.780 & -0.100 \\
    11.0&  1.0 &   5.324 &  -0.062 &   6.643 &   5.163 &   5.472 & -0.080 \\
    11.0&  3.0 &   5.329 &   0.047 &   6.650 &   5.182 &   5.491 & -0.090 \\
    11.0& 30.0 &   5.380 &   0.182 &   6.751 &   5.249 &   5.598 & -0.090 \\
\hline
\end{tabular}\\
\vspace{0.20cm}\\
\noindent
\begin{tabular}{rr|rrrrrr}
\multicolumn{7}{c}{\normalsize{\hspace*{3.5cm}$^{  61}_{32}$ GE $_{ 29} \rightarrow \hspace*{0.2cm}^{  61}_{33}$ AS $_{ 28}$}}\\
\hline
Aden& T9&  E+Cap &   E-Dec &  ANuEn &  GamEn &  ProEn &  PPEm \\
\hline
     3.0&  1.0 & -84.099 &    -100 & -84.048 &    -100 & -86.579 &  0.000 \\
     3.0&  3.0 & -28.886 &    -100 & -28.734 &    -100 & -29.943 &  0.000 \\
     3.0& 30.0 &  -1.138 &    -100 &  -0.171 &    -100 &  -0.422 &  0.000 \\
     7.0&  1.0 & -89.919 &    -100 & -89.868 &    -100 & -92.399 &  0.000 \\
     7.0&  3.0 & -30.601 &    -100 & -30.450 &    -100 & -31.659 &  0.000 \\
     7.0& 30.0 &  -1.142 &    -100 &  -0.174 &    -100 &  -0.425 &  0.000 \\
    11.0&  1.0 &    -100 &    -100 &    -100 &    -100 &    -100 &-100\\
    11.0&  3.0 & -69.079 &    -100 & -68.927 &    -100 & -70.136 &  0.000 \\
    11.0& 30.0 &  -5.005 &    -100 &  -4.037 &    -100 &  -4.289 &  0.000 \\
\hline
\end{tabular}\\
\vspace{0.20cm}\\
\begin{tabular}{rr|rrrrrr}
\multicolumn{7}{c}{\normalsize{\hspace*{3.5cm}$^{  61}_{32}$ GE $_{ 29} \rightarrow \hspace*{0.2cm}^{  61}_{31}$ GA $_{ 30}$}}\\
\hline
Aden& T9&  E-Cap &   E+Dec &  NuEn &  GamEn &  ProEn &  PPEm \\
\hline
     3.0&  1.0 &  -3.734 &   0.908 &   1.606 &  -0.422 &   1.150 & -0.050 \\
     3.0&  3.0 &  -1.486 &   0.897 &   1.605 &  -0.220 &   1.135 & -0.090 \\
     3.0& 30.0 &   2.525 &   0.878 &   3.690 &   0.635 &   3.534 & -0.020 \\
     7.0&  1.0 &  -0.068 &   0.908 &   1.689 &  -0.405 &   1.217 & -0.050 \\
     7.0&  3.0 &  -0.080 &   0.897 &   1.685 &  -0.204 &   1.203 & -0.090 \\
     7.0& 30.0 &   2.529 &   0.878 &   3.694 &   0.638 &   3.537 & -0.020 \\
    11.0&  1.0 &   5.204 &   0.908 &   6.509 &   2.667 &   6.257 & -0.003 \\
    11.0&  3.0 &   5.208 &   0.897 &   6.512 &   2.864 &   6.264 & -0.005 \\
    11.0& 30.0 &   5.290 &   0.943 &   6.642 &   2.990 &   6.374 & -0.006 \\
\hline
\end{tabular}\\
\vspace{0.20cm}\\
\noindent
\begin{tabular}{rr|rrrrrr}
\multicolumn{7}{c}{\normalsize{\hspace*{3.5cm}$^{  62}_{24}$ CR $_{ 38} \rightarrow \hspace*{0.2cm}^{  62}_{25}$ MN $_{ 37}$}}\\
\hline
Aden & T9& E+Cap &   E-Dec &  ANuEn &  GamEn &  NeuEn &  PNEm \\
\hline
     3.0&  1.0 &  -5.392 &   0.763 &   1.310 &   0.641 &  -3.109 & -3.000 \\
     3.0&  3.0 &  -2.323 &   0.762 &   1.310 &   0.641 &  -2.852 & -3.000 \\
     3.0& 30.0 &   2.406 &   1.325 &   3.554 &   2.242 &   3.252 & -0.100 \\
     7.0&  1.0 & -11.212 &   0.746 &   1.283 &   0.618 &  -3.746 & -4.000 \\
     7.0&  3.0 &  -4.028 &   0.748 &   1.287 &   0.621 &  -3.439 & -3.000 \\
     7.0& 30.0 &   2.403 &   1.324 &   3.551 &   2.239 &   3.249 & -0.100 \\
    11.0&  1.0 &    -100 & -84.295 & -84.889 & -85.303 &    -100 &-30.000 \\
    11.0&  3.0 & -42.506 & -28.191 & -28.322 & -28.847 & -38.565 &-10.000 \\
    11.0& 30.0 &  -1.448 &  -1.522 &  -0.223 &  -1.209 &  -0.606 & -0.300 \\
\hline
\end{tabular}\\
\vspace{0.20cm}\\
\begin{tabular}{rr|rrrrrr}
\multicolumn{7}{c}{\normalsize{\hspace*{3.5cm}$^{  62}_{24}$ CR $_{ 38} \rightarrow \hspace*{0.2cm}^{  62}_{23}$ V  $_{ 39}$}}\\
\hline
Aden & T9& E-Cap &   E+Dec &  NuEn &  GamEn &  NeuEn &  PNEm \\
\hline
     3.0&  1.0 & -86.460 &    -100 & -87.043 & -87.491 &    -100 &-20.000 \\
     3.0&  3.0 & -28.766 &    -100 & -28.862 & -29.441 & -35.420 & -6.000 \\
     3.0& 30.0 &  -0.349 &    -100 &   0.615 &  -0.592 &  -0.457 & -0.400 \\
     7.0&  1.0 & -80.640 &    -100 & -81.223 & -81.671 &    -100 &-20.000 \\
     7.0&  3.0 & -27.050 &    -100 & -27.146 & -27.725 & -33.704 & -6.000 \\
     7.0& 30.0 &  -0.345 &    -100 &   0.618 &  -0.588 &  -0.453 & -0.400 \\
    11.0&  1.0 &   2.457 &    -100 &   3.165 &   2.253 &  -0.547 & -3.000 \\
    11.0&  3.0 &   2.463 &    -100 &   3.178 &   2.260 &  -0.455 & -3.000 \\
    11.0& 30.0 &   3.268 &    -100 &   4.328 &   2.984 &   3.252 & -0.300 \\
\hline
\end{tabular}\\
\vspace{0.20cm}\\
\noindent
\begin{tabular}{rr|rrrrrr}
\multicolumn{7}{c}{\normalsize{\hspace*{3.5cm}$^{  62}_{25}$ MN $_{ 37} \rightarrow \hspace*{0.2cm}^{  62}_{26}$ FE $_{ 36}$}}\\
\hline
Aden & T9& E+Cap &   E-Dec &  ANuEn &  GamEn &  NeuEn &  PNEm \\
\hline
     3.0&  1.0 &  -4.326 &   2.045 &   2.619 &   2.597 &  -4.421 & -6.000 \\
     3.0&  3.0 &  -1.211 &   2.118 &   2.692 &   2.674 &  -3.589 & -5.000 \\
     3.0& 30.0 &   2.637 &   1.943 &   3.900 &   3.327 &   2.675 & -0.500 \\
     7.0&  1.0 & -10.146 &   2.032 &   2.596 &   2.584 &  -5.031 & -6.000 \\
     7.0&  3.0 &  -2.916 &   2.107 &   2.673 &   2.663 &  -4.049 & -6.000 \\
     7.0& 30.0 &   2.634 &   1.942 &   3.897 &   3.325 &   2.671 & -0.500 \\
    11.0&  1.0 &    -100 & -75.143 & -75.738 & -75.201 &    -100 &-40.000 \\
    11.0&  3.0 & -41.393 & -27.369 & -27.503 & -27.227 & -37.684 & -9.000 \\
    11.0& 30.0 &  -1.207 &  -1.121 &   0.117 &  -0.221 &  -1.171 & -0.600 \\
\hline
\end{tabular}\\
\vspace{0.20cm}\\
\begin{tabular}{rr|rrrrrr}
\multicolumn{7}{c}{\normalsize{\hspace*{3.5cm}$^{  62}_{25}$ MN $_{ 37} \rightarrow \hspace*{0.2cm}^{  62}_{24}$ CR $_{ 38}$}}\\
\hline
Aden & T9& E-Cap &   E+Dec &  NuEn &  GamEn &  NeuEn &  PNEm \\
\hline
     3.0&  1.0 & -41.507 &    -100 & -42.085 & -42.606 & -80.466 &-40.000 \\
     3.0&  3.0 & -13.928 &    -100 & -14.013 & -14.584 & -26.643 &-10.000 \\
     3.0& 30.0 &   1.960 &    -100 &   2.961 &   2.449 &   1.225 & -1.000 \\
     7.0&  1.0 & -35.687 &    -100 & -36.265 & -36.786 & -74.646 &-40.000 \\
     7.0&  3.0 & -12.212 &    -100 & -12.297 & -12.868 & -24.927 &-10.000 \\
     7.0& 30.0 &   1.963 &    -100 &   2.964 &   2.453 &   1.228 & -1.000 \\
    11.0&  1.0 &   4.675 &    -100 &   5.666 &   5.246 &   2.597 & -2.000 \\
    11.0&  3.0 &   4.802 &    -100 &   5.818 &   5.301 &   2.717 & -2.000 \\
    11.0& 30.0 &   5.212 &    -100 &   6.404 &   5.686 &   4.688 & -0.800 \\
\hline
\end{tabular}\\
\vspace{0.20cm}\\
\noindent
\begin{tabular}{rr|rrrrrr}
\multicolumn{7}{c}{\normalsize{\hspace*{3.5cm}$^{  62}_{26}$ FE $_{ 36} \rightarrow \hspace*{0.2cm}^{  62}_{27}$ CO $_{ 35}$}}\\
\hline
Aden & T9& E+Cap &   E-Dec &  ANuEn &  GamEn &  NeuEn &  PNEm \\
\hline
     3.0&  1.0 &  -6.785 &  -1.983 &  -1.906 &  -2.270 & -31.626 &-30.000 \\
     3.0&  3.0 &  -3.674 &  -2.004 &  -1.902 &  -2.279 & -11.618 & -9.000 \\
     3.0& 30.0 &   1.710 &   0.448 &   2.894 &   1.991 &   1.941 & -0.600 \\
     7.0&  1.0 & -12.605 &  -2.211 &  -2.247 &  -2.504 & -35.924 &-30.000 \\
     7.0&  3.0 &  -5.380 &  -2.161 &  -2.142 &  -2.450 & -13.273 &-10.000 \\
     7.0& 30.0 &   1.707 &   0.447 &   2.891 &   1.988 &   1.937 & -0.600 \\
    11.0&  1.0 &    -100 &    -100 &    -100 &    -100 &    -100 &-100\\
    11.0&  3.0 & -43.857 & -36.666 & -36.811 & -37.055 & -51.396 &-10.000 \\
    11.0& 30.0 &  -2.139 &  -2.707 &  -0.945 &  -1.785 &  -1.923 & -0.600 \\
\hline
\end{tabular}\\
\vspace{0.20cm}\\
\begin{tabular}{rr|rrrrrr}
\multicolumn{7}{c}{\normalsize{\hspace*{3.5cm}$^{  62}_{26}$ FE $_{ 36} \rightarrow \hspace*{0.2cm}^{  62}_{25}$ MN $_{ 37}$}}\\
\hline
Aden & T9& E-Cap &   E+Dec &  NuEn &  GamEn &  NeuEn &  PNEm \\
\hline
     3.0&  1.0 & -58.245 &    -100 & -58.824 & -58.843 & -87.130 &-30.000 \\
     3.0&  3.0 & -18.997 &    -100 & -19.082 & -19.257 & -28.983 & -9.000 \\
     3.0& 30.0 &   1.600 &    -100 &   2.602 &   1.949 &   0.067 & -2.000 \\
     7.0&  1.0 & -52.426 &    -100 & -53.004 & -53.023 & -81.310 &-30.000 \\
     7.0&  3.0 & -17.281 &    -100 & -17.367 & -17.542 & -27.268 & -9.000 \\
     7.0& 30.0 &   1.603 &    -100 &   2.606 &   1.953 &   0.071 & -2.000 \\
    11.0&  1.0 &   2.736 &    -100 &   3.644 &   3.104 &   1.368 & -1.000 \\
    11.0&  3.0 &   2.726 &    -100 &   3.636 &   3.093 &   1.369 & -1.000 \\
    11.0& 30.0 &   4.813 &    -100 &   6.020 &   5.161 &   3.615 & -1.000 \\
\hline
\end{tabular}\\
\vspace{0.20cm}\\
\noindent
\begin{tabular}{rr|rrrrrr}
\multicolumn{7}{c}{\normalsize{\hspace*{3.5cm}$^{  62}_{27}$ CO $_{ 35} \rightarrow \hspace*{0.2cm}^{  62}_{28}$ NI $_{ 34}$}}\\
\hline
Aden & T9& E+Cap &   E-Dec &  ANuEn &  GamEn &  NeuEn &  PNEm \\
\hline
     3.0&  1.0 &  -7.049 &  -2.252 &  -1.970 &  -1.963 & -35.987 &-30.000 \\
     3.0&  3.0 &  -3.426 &  -1.930 &  -1.706 &  -1.428 & -12.263 &-10.000 \\
     3.0& 30.0 &   2.405 &  -0.320 &   3.505 &   3.047 &   2.769 & -0.300 \\
     7.0&  1.0 & -12.869 &  -2.359 &  -2.088 &  -2.133 & -40.319 &-40.000 \\
     7.0&  3.0 &  -5.134 &  -2.034 &  -1.852 &  -1.571 & -13.943 &-10.000 \\
     7.0& 30.0 &   2.401 &  -0.321 &   3.502 &   3.044 &   2.765 & -0.300 \\
    11.0&  1.0 &    -100 &    -100 &    -100 &    -100 &    -100 &-100\\
    11.0&  3.0 & -43.611 & -37.112 & -37.278 & -36.937 & -52.392 &-10.000 \\
    11.0& 30.0 &  -1.452 &  -3.546 &  -0.355 &  -0.802 &  -1.095 & -0.300 \\
\hline
\end{tabular}\\
\vspace{0.20cm}\\
\begin{tabular}{rr|rrrrrr}
\multicolumn{7}{c}{\normalsize{\hspace*{3.5cm}$^{  62}_{27}$ CO $_{ 35} \rightarrow \hspace*{0.2cm}^{  62}_{26}$ FE $_{ 36}$}}\\
\hline
Aden & T9& E-Cap &   E+Dec &  NuEn &  GamEn &  NeuEn &  PNEm \\
\hline
     3.0&  1.0 & -19.857 &    -100 & -20.418 & -34.016 & -63.620 &-40.000 \\
     3.0&  3.0 &  -8.449 &    -100 &  -8.495 & -11.661 & -20.831 &-10.000 \\
     3.0& 30.0 &   1.978 &    -100 &   2.985 &   2.426 &   2.100 & -0.200 \\
     7.0&  1.0 & -14.037 &    -100 & -14.599 & -28.197 & -57.800 &-40.000 \\
     7.0&  3.0 &  -6.733 &    -100 &  -6.779 &  -9.947 & -19.116 &-10.000 \\
     7.0& 30.0 &   1.982 &    -100 &   2.989 &   2.430 &   2.104 & -0.200 \\
    11.0&  1.0 &   4.480 &    -100 &   5.576 &   5.196 &   2.418 & -2.000 \\
    11.0&  3.0 &   4.548 &    -100 &   5.645 &   5.277 &   2.648 & -2.000 \\
    11.0& 30.0 &   5.180 &    -100 &   6.388 &   5.582 &   5.365 & -0.200 \\
\hline
\end{tabular}\\
\vspace{0.20cm}\\
\noindent
\begin{tabular}{rr|rrrrrr}
\multicolumn{7}{c}{\normalsize{\hspace*{3.5cm}$^{  62}_{28}$ NI $_{ 34} \rightarrow \hspace*{0.2cm}^{  62}_{29}$ CU $_{ 33}$}}\\
\hline
Aden& T9&  E+Cap &   E-Dec &  ANuEn &  GamEn &  ProEn &  PPEm \\
\hline
     3.0&  1.0 & -25.123 & -27.480 & -25.433 & -27.830 & -59.874 &-30.000 \\
     3.0&  3.0 & -10.330 & -11.506 & -10.154 & -11.067 & -19.937 & -9.000 \\
     3.0& 30.0 &   1.075 &  -2.245 &   2.123 &   1.243 &   1.684 & -0.200 \\
     7.0&  1.0 & -30.943 & -29.687 & -30.089 & -31.091 & -65.639 &-30.000 \\
     7.0&  3.0 & -12.044 & -11.837 & -11.632 & -12.355 & -21.652 & -9.000 \\
     7.0& 30.0 &   1.072 &  -2.246 &   2.120 &   1.239 &   1.680 & -0.200 \\
    11.0&  1.0 &    -100 &    -100 &    -100 &    -100 &    -100 &-100\\
    11.0&  3.0 & -50.522 & -48.955 & -49.128 & -49.321 & -60.130 &-10.000 \\
    11.0& 30.0 &  -2.787 &  -5.618 &  -1.740 &  -2.616 &  -2.182 & -0.200 \\
\hline
\end{tabular}\\
\vspace{0.20cm}\\
\begin{tabular}{rr|rrrrrr}
\multicolumn{7}{c}{\normalsize{\hspace*{3.5cm}$^{  62}_{28}$ NI $_{ 34} \rightarrow \hspace*{0.2cm}^{  62}_{27}$ CO $_{ 35}$}}\\
\hline
Aden & T9& E-Cap &   E+Dec &  NuEn &  GamEn &  NeuEn &  PNEm \\
\hline
     3.0&  1.0 & -33.959 & -39.914 & -34.525 & -36.086 & -70.629 &-40.000 \\
     3.0&  3.0 & -12.726 & -16.425 & -12.772 & -13.079 & -23.151 &-10.000 \\
     3.0& 30.0 &   1.476 &  -5.461 &   2.488 &   1.816 &   1.592 & -0.300 \\
     7.0&  1.0 & -28.153 & -39.914 & -28.719 & -30.300 & -64.809 &-40.000 \\
     7.0&  3.0 & -11.017 & -16.407 & -11.063 & -11.368 & -21.435 &-10.000 \\
     7.0& 30.0 &   1.479 &  -5.459 &   2.492 &   1.819 &   1.595 & -0.300 \\
    11.0&  1.0 &   4.341 & -39.914 &   5.435 &   4.744 &   1.895 & -3.000 \\
    11.0&  3.0 &   4.338 & -16.407 &   5.433 &   4.740 &   1.899 & -3.000 \\
    11.0& 30.0 &   4.659 &  -5.233 &   5.871 &   4.981 &   4.826 & -0.200 \\
\hline
\end{tabular}\\
\vspace{0.20cm}\\
\noindent
\begin{tabular}{rr|rrrrrr}
\multicolumn{7}{c}{\normalsize{\hspace*{3.5cm}$^{  62}_{29}$ CU $_{ 33} \rightarrow \hspace*{0.2cm}^{  62}_{30}$ ZN $_{ 32}$}}\\
\hline
Aden& T9&  E+Cap &   E-Dec &  ANuEn &  GamEn &  ProEn &  PPEm \\
\hline
     3.0&  1.0 & -12.898 &    -100 & -13.312 & -28.251 & -50.315 &-40.000 \\
     3.0&  3.0 &  -6.753 &    -100 &  -6.624 &  -9.546 & -16.325 & -9.000 \\
     3.0& 30.0 &   2.015 &    -100 &   3.054 &   2.303 &   2.370 & -0.200 \\
     7.0&  1.0 & -18.718 &    -100 & -19.132 & -34.071 & -56.135 &-40.000 \\
     7.0&  3.0 &  -8.467 &    -100 &  -8.339 & -11.261 & -18.041 & -9.000 \\
     7.0& 30.0 &   2.011 &    -100 &   3.051 &   2.299 &   2.367 & -0.200 \\
    11.0&  1.0 &    -100 &    -100 &    -100 &    -100 &    -100 &-100\\
    11.0&  3.0 & -46.945 &    -100 & -46.816 & -49.738 & -56.518 & -9.000 \\
    11.0& 30.0 &  -1.849 &    -100 &  -0.811 &  -1.560 &  -1.495 & -0.200 \\
\hline
\end{tabular}\\
\vspace{0.20cm}\\
\begin{tabular}{rr|rrrrrr}
\multicolumn{7}{c}{\normalsize{\hspace*{3.5cm}$^{  62}_{29}$ CU $_{ 33} \rightarrow \hspace*{0.2cm}^{  62}_{28}$ NI $_{ 34}$}}\\
\hline
Aden & T9& E-Cap &   E+Dec &  NuEn &  GamEn &  NeuEn &  PNEm \\
\hline
     3.0&  1.0 &  -6.339 &  -3.108 &  -2.901 &  -5.486 & -44.287 &-40.000 \\
     3.0&  3.0 &  -3.924 &  -3.328 &  -2.958 &  -3.491 & -14.834 &-10.000 \\
     3.0& 30.0 &   2.760 &  -1.474 &   3.857 &   3.540 &   2.396 & -0.800 \\
     7.0&  1.0 &  -2.605 &  -3.108 &  -1.938 &  -3.088 & -38.467 &-30.000 \\
     7.0&  3.0 &  -2.427 &  -3.326 &  -1.928 &  -2.011 & -13.119 &-10.000 \\
     7.0& 30.0 &   2.763 &  -1.473 &   3.860 &   3.543 &   2.400 & -0.800 \\
    11.0&  1.0 &   4.902 &  -3.108 &   6.113 &   5.675 &   3.941 & -0.900 \\
    11.0&  3.0 &   4.906 &  -3.326 &   6.116 &   5.684 &   4.029 & -0.900 \\
    11.0& 30.0 &   5.612 &  -1.302 &   6.928 &   6.339 &   5.556 & -0.500 \\
\hline
\end{tabular}\\
\vspace{0.20cm}\\
\noindent
\begin{tabular}{rr|rrrrrr}
\multicolumn{7}{c}{\normalsize{\hspace*{3.5cm}$^{  62}_{30}$ ZN $_{ 32} \rightarrow \hspace*{0.2cm}^{  62}_{31}$ GA $_{ 31}$}}\\
\hline
Aden& T9&  E+Cap &   E-Dec &  ANuEn &  GamEn &  ProEn &  PPEm \\
\hline
     3.0&  1.0 & -50.397 &    -100 & -50.346 & -53.579 & -71.594 &-20.000 \\
     3.0&  3.0 & -16.950 &    -100 & -16.794 & -18.149 & -23.695 & -6.000 \\
     3.0& 30.0 &   0.985 &    -100 &   1.993 &   0.743 &   1.549 & -0.100 \\
     7.0&  1.0 & -56.217 &    -100 & -56.166 & -59.399 & -77.414 &-20.000 \\
     7.0&  3.0 & -18.666 &    -100 & -18.510 & -19.864 & -25.411 & -6.000 \\
     7.0& 30.0 &   0.981 &    -100 &   1.990 &   0.739 &   1.545 & -0.100 \\
    11.0&  1.0 &    -100 &    -100 &    -100 &    -100 &    -100 &-100\\
    11.0&  3.0 & -57.143 &    -100 & -56.987 & -58.342 & -63.888 & -6.000 \\
    11.0& 30.0 &  -2.881 &    -100 &  -1.873 &  -3.121 &  -2.318 & -0.100 \\
\hline
\end{tabular}\\
\vspace{0.20cm}\\
\begin{tabular}{rr|rrrrrr}
\multicolumn{7}{c}{\normalsize{\hspace*{3.5cm}$^{  62}_{30}$ ZN $_{ 32} \rightarrow \hspace*{0.2cm}^{  62}_{29}$ CU $_{ 33}$}}\\
\hline
Aden& T9&  E-Cap &   E+Dec &  NuEn &  GamEn &  ProEn &  PPEm \\
\hline
     3.0&  1.0 &  -6.223 &  -5.757 &  -5.863 &  -6.530 & -34.136 &-30.000 \\
     3.0&  3.0 &  -3.681 &  -5.739 &  -3.436 &  -3.857 & -11.806 & -8.000 \\
     3.0& 30.0 &   2.464 &  -0.657 &   3.550 &   2.504 &   2.773 & -0.200 \\
     7.0&  1.0 &  -2.326 &  -5.757 &  -2.120 &  -2.564 & -28.339 &-30.000 \\
     7.0&  3.0 &  -2.165 &  -5.718 &  -1.884 &  -2.299 & -10.095 & -8.000 \\
     7.0& 30.0 &   2.467 &  -0.656 &   3.553 &   2.508 &   2.777 & -0.200 \\
    11.0&  1.0 &   4.997 &  -5.757 &   6.222 &   5.561 &   4.429 & -1.000 \\
    11.0&  3.0 &   4.998 &  -5.718 &   6.223 &   5.562 &   4.430 & -1.000 \\
    11.0& 30.0 &   5.347 &  -0.531 &   6.654 &   5.284 &   5.775 & -0.100 \\
\hline
\end{tabular}\\
\vspace{0.20cm}\\
\noindent
\begin{tabular}{rr|rrrrrr}
\multicolumn{7}{c}{\normalsize{\hspace*{3.5cm}$^{  62}_{31}$ GA $_{ 31} \rightarrow \hspace*{0.2cm}^{  62}_{32}$ GE $_{ 30}$}}\\
\hline
Aden& T9&  E+Cap &   E-Dec &  ANuEn &  GamEn &  ProEn &  PPEm \\
\hline
     3.0&  1.0 & -70.479 &    -100 & -70.428 &    -100 & -71.344 &  0.000 \\
     3.0&  3.0 & -23.563 &    -100 & -23.410 &    -100 & -23.800 &  0.000 \\
     3.0& 30.0 &   0.256 &    -100 &   1.235 &    -100 &   1.038 &  0.000 \\
     7.0&  1.0 & -76.298 &    -100 & -76.248 &    -100 & -77.164 &  0.000 \\
     7.0&  3.0 & -25.278 &    -100 & -25.126 &    -100 & -25.516 &  0.000 \\
     7.0& 30.0 &   0.252 &    -100 &   1.232 &    -100 &   1.035 &  0.000 \\
    11.0&  1.0 &    -100 &    -100 &    -100 &    -100 &    -100 &-100\\
    11.0&  3.0 & -63.756 &    -100 & -63.603 &    -100 & -63.993 &  0.000 \\
    11.0& 30.0 &  -3.611 &    -100 &  -2.631 &    -100 &  -2.828 &  0.000 \\
\hline
\end{tabular}\\
\vspace{0.20cm}\\
\begin{tabular}{rr|rrrrrr}
\multicolumn{7}{c}{\normalsize{\hspace*{3.5cm}$^{  62}_{31}$ GA $_{ 31} \rightarrow \hspace*{0.2cm}^{  62}_{30}$ ZN $_{ 32}$}}\\
\hline
Aden& T9&  E-Cap &   E+Dec &  NuEn &  GamEn &  ProEn &  PPEm \\
\hline
     3.0&  1.0 &  -3.747 &   0.732 &   1.358 &  -0.182 &  -5.909 & -6.000 \\
     3.0&  3.0 &  -1.564 &   0.545 &   1.169 &   0.057 &  -3.215 & -4.000 \\
     3.0& 30.0 &   2.758 &   0.040 &   3.869 &   2.853 &   3.334 & -0.100 \\
     7.0&  1.0 &  -0.070 &   0.732 &   1.480 &   0.144 &  -2.297 & -3.000 \\
     7.0&  3.0 &  -0.150 &   0.545 &   1.308 &   0.352 &  -1.685 & -2.000 \\
     7.0& 30.0 &   2.761 &   0.041 &   3.872 &   2.856 &   3.337 & -0.100 \\
    11.0&  1.0 &   5.277 &   0.732 &   6.576 &   5.130 &   5.704 & -0.200 \\
    11.0&  3.0 &   5.342 &   0.545 &   6.627 &   5.200 &   5.830 & -0.100 \\
    11.0& 30.0 &   5.574 &   0.147 &   6.904 &   5.455 &   6.250 & -0.070 \\
\hline
\end{tabular}\\
\vspace{0.20cm}\\
\noindent
\begin{tabular}{rr|rrrrrr}
\multicolumn{7}{c}{\normalsize{\hspace*{3.5cm}$^{  62}_{32}$ GE $_{ 30} \rightarrow \hspace*{0.2cm}^{  62}_{33}$ AS $_{ 29}$}}\\
\hline
Aden& T9&  E+Cap &   E-Dec &  ANuEn &  GamEn &  ProEn &  PPEm \\
\hline
     3.0&  1.0 & -94.223 &    -100 & -94.173 &    -100 & -94.981 &  0.000 \\
     3.0&  3.0 & -32.750 &    -100 & -32.599 &    -100 & -33.045 &  0.000 \\
     3.0& 30.0 &  -1.084 &    -100 &  -0.121 &    -100 &  -0.451 &  0.000 \\
     7.0&  1.0 &    -100 &    -100 & -99.992 &    -100 &    -100 &-100\\
     7.0&  3.0 & -34.465 &    -100 & -34.315 &    -100 & -34.761 &  0.000 \\
     7.0& 30.0 &  -1.088 &    -100 &  -0.124 &    -100 &  -0.455 &  0.000 \\
    11.0&  1.0 &    -100 &    -100 &    -100 &    -100 &    -100 &-100\\
    11.0&  3.0 & -72.943 &    -100 & -72.792 &    -100 & -73.238 &  0.000 \\
    11.0& 30.0 &  -4.951 &    -100 &  -3.988 &    -100 &  -4.318 &  0.000 \\
\hline
\end{tabular}\\
\vspace{0.20cm}\\
\begin{tabular}{rr|rrrrrr}
\multicolumn{7}{c}{\normalsize{\hspace*{3.5cm}$^{  62}_{32}$ GE $_{ 30} \rightarrow \hspace*{0.2cm}^{  62}_{31}$ GA $_{ 31}$}}\\
\hline
Aden& T9&  E-Cap &   E+Dec &  NuEn &  GamEn &  ProEn &  PPEm \\
\hline
     3.0&  1.0 &  -4.333 &  -0.077 &   0.534 &  -0.158 &  -1.797 & -1.000 \\
     3.0&  3.0 &  -1.999 &  -0.077 &   0.542 &  -0.154 &  -1.508 & -1.000 \\
     3.0& 30.0 &   2.419 &  -0.165 &   3.536 &   1.571 &   3.108 & -0.040 \\
     7.0&  1.0 &  -0.610 &  -0.077 &   0.697 &  -0.054 &  -0.439 & -0.900 \\
     7.0&  3.0 &  -0.557 &  -0.077 &   0.710 &  -0.052 &  -0.294 & -0.800 \\
     7.0& 30.0 &   2.423 &  -0.165 &   3.540 &   1.574 &   3.111 & -0.040 \\
    11.0&  1.0 &   5.130 &  -0.077 &   6.426 &   4.085 &   5.863 & -0.030 \\
    11.0&  3.0 &   5.130 &  -0.077 &   6.427 &   4.085 &   5.864 & -0.020 \\
    11.0& 30.0 &   5.186 &  -0.077 &   6.533 &   4.120 &   5.926 & -0.020 \\
\hline
\end{tabular}\\
\vspace{0.20cm}\\
\noindent
\begin{tabular}{rr|rrrrrr}
\multicolumn{7}{c}{\normalsize{\hspace*{3.5cm}$^{  63}_{24}$ CR $_{ 39} \rightarrow \hspace*{0.2cm}^{  63}_{25}$ MN $_{ 38}$}}\\
\hline
Aden & T9& E+Cap &   E-Dec &  ANuEn &  GamEn &  NeuEn &  PNEm \\
\hline
     3.0&  1.0 &  -5.334 &   0.912 &   1.553 &   1.364 &  -0.566 & -1.000 \\
     3.0&  3.0 &  -2.268 &   0.986 &   1.661 &   1.336 &  -0.524 & -2.000 \\
     3.0& 30.0 &   2.213 &   1.007 &   3.348 &   2.131 &   3.015 & -0.090 \\
     7.0&  1.0 & -11.154 &   0.899 &   1.535 &   1.351 &  -0.686 & -2.000 \\
     7.0&  3.0 &  -3.972 &   0.976 &   1.648 &   1.324 &  -0.619 & -2.000 \\
     7.0& 30.0 &   2.209 &   1.006 &   3.345 &   2.128 &   3.012 & -0.090 \\
    11.0&  1.0 &    -100 & -65.011 & -65.604 & -66.731 &    -100 &-40.000 \\
    11.0&  3.0 & -42.450 & -22.777 & -22.905 & -24.226 & -36.400 &-10.000 \\
    11.0& 30.0 &  -1.643 &  -1.799 &  -0.438 &  -1.344 &  -0.846 & -0.300 \\
\hline
\end{tabular}\\
\vspace{0.20cm}\\
\begin{tabular}{rr|rrrrrr}
\multicolumn{7}{c}{\normalsize{\hspace*{3.5cm}$^{  63}_{24}$ CR $_{ 39} \rightarrow \hspace*{0.2cm}^{  63}_{23}$ V  $_{ 40}$}}\\
\hline
Aden & T9& E-Cap &   E+Dec &  NuEn &  GamEn &  NeuEn &  PNEm \\
\hline
     3.0&  1.0 & -76.041 &    -100 & -76.624 & -77.463 &    -100 &-30.000 \\
     3.0&  3.0 & -26.468 &    -100 & -26.565 & -27.314 & -36.182 & -9.000 \\
     3.0& 30.0 &  -1.047 &    -100 &  -0.093 &  -0.711 &  -1.437 & -0.600 \\
     7.0&  1.0 & -70.221 &    -100 & -70.804 & -71.643 & -99.671 &-30.000 \\
     7.0&  3.0 & -24.753 &    -100 & -24.850 & -25.598 & -34.466 & -9.000 \\
     7.0& 30.0 &  -1.043 &    -100 &  -0.089 &  -0.708 &  -1.434 & -0.600 \\
    11.0&  1.0 &   1.982 &    -100 &   2.644 &   2.497 &   0.607 & -1.000 \\
    11.0&  3.0 &   2.085 &    -100 &   2.818 &   2.478 &   0.727 & -1.000 \\
    11.0& 30.0 &   2.654 &    -100 &   3.675 &   3.012 &   2.380 & -0.500 \\
\hline
\end{tabular}\\
\vspace{0.20cm}\\
\noindent
\begin{tabular}{rr|rrrrrr}
\multicolumn{7}{c}{\normalsize{\hspace*{3.5cm}$^{  63}_{25}$ MN $_{ 38} \rightarrow \hspace*{0.2cm}^{  63}_{26}$ FE $_{ 37}$}}\\
\hline
Aden & T9& E+Cap &   E-Dec &  ANuEn &  GamEn &  NeuEn &  PNEm \\
\hline
     3.0&  1.0 &  -5.735 &   0.784 &   1.434 &   0.604 &  -2.318 & -3.000 \\
     3.0&  3.0 &  -2.758 &   0.686 &   1.336 &   0.567 &  -2.224 & -3.000 \\
     3.0& 30.0 &   2.579 &   2.160 &   3.878 &   2.993 &   2.966 & -0.200 \\
     7.0&  1.0 & -11.555 &   0.775 &   1.419 &   0.589 &  -2.539 & -3.000 \\
     7.0&  3.0 &  -4.462 &   0.678 &   1.323 &   0.554 &  -2.393 & -3.000 \\
     7.0& 30.0 &   2.576 &   2.160 &   3.875 &   2.990 &   2.962 & -0.200 \\
    11.0&  1.0 &    -100 & -75.724 & -76.318 & -77.149 &    -100 &-30.000 \\
    11.0&  3.0 & -42.939 & -25.914 & -26.045 & -26.656 & -34.802 & -8.000 \\
    11.0& 30.0 &  -1.265 &  -0.750 &   0.208 &  -0.262 &  -0.726 & -0.400 \\
\hline
\end{tabular}\\
\vspace{0.20cm}\\
\begin{tabular}{rr|rrrrrr}
\multicolumn{7}{c}{\normalsize{\hspace*{3.5cm}$^{  63}_{25}$ MN $_{ 38} \rightarrow \hspace*{0.2cm}^{  63}_{24}$ CR $_{ 39}$}}\\
\hline
Aden & T9& E-Cap &   E+Dec &  NuEn &  GamEn &  NeuEn &  PNEm \\
\hline
     3.0&  1.0 & -60.409 &    -100 & -60.988 & -61.700 & -79.035 &-20.000 \\
     3.0&  3.0 & -19.517 &    -100 & -19.602 & -20.092 & -26.365 & -6.000 \\
     3.0& 30.0 &   1.916 &    -100 &   2.919 &   2.045 &   1.258 & -1.000 \\
     7.0&  1.0 & -54.589 &    -100 & -55.168 & -55.881 & -73.215 &-20.000 \\
     7.0&  3.0 & -17.802 &    -100 & -17.887 & -18.376 & -24.649 & -6.000 \\
     7.0& 30.0 &   1.919 &    -100 &   2.923 &   2.049 &   1.262 & -1.000 \\
    11.0&  1.0 &   3.653 &    -100 &   4.561 &   3.871 &   2.422 & -1.000 \\
    11.0&  3.0 &   3.508 &    -100 &   4.416 &   3.719 &   2.437 & -1.000 \\
    11.0& 30.0 &   5.142 &    -100 &   6.342 &   5.242 &   4.764 & -0.800 \\
\hline
\end{tabular}\\
\vspace{0.20cm}\\
\noindent
\begin{tabular}{rr|rrrrrr}
\multicolumn{7}{c}{\normalsize{\hspace*{3.5cm}$^{  63}_{26}$ FE $_{ 37} \rightarrow \hspace*{0.2cm}^{  63}_{27}$ CO $_{ 36}$}}\\
\hline
Aden & T9& E+Cap &   E-Dec &  ANuEn &  GamEn &  NeuEn &  PNEm \\
\hline
     3.0&  1.0 &  -6.883 &  -0.945 &  -0.447 &  -1.448 & -24.067 &-20.000 \\
     3.0&  3.0 &  -3.665 &  -0.764 &  -0.251 &  -1.305 &  -9.370 & -8.000 \\
     3.0& 30.0 &   1.685 &   0.359 &   2.858 &   2.301 &   1.888 & -0.500 \\
     7.0&  1.0 & -12.703 &  -0.969 &  -0.484 &  -1.530 & -27.257 &-30.000 \\
     7.0&  3.0 &  -5.371 &  -0.783 &  -0.280 &  -1.371 & -10.622 & -9.000 \\
     7.0& 30.0 &   1.682 &   0.359 &   2.855 &   2.298 &   1.885 & -0.500 \\
    11.0&  1.0 &    -100 & -91.565 & -92.162 & -93.307 &    -100 &-50.000 \\
    11.0&  3.0 & -43.848 & -31.933 & -32.072 & -33.151 & -48.505 &-20.000 \\
    11.0& 30.0 &  -2.165 &  -2.786 &  -0.983 &  -1.463 &  -1.974 & -0.500 \\
\hline
\end{tabular}\\
\vspace{0.20cm}\\
\begin{tabular}{rr|rrrrrr}
\multicolumn{7}{c}{\normalsize{\hspace*{3.5cm}$^{  63}_{26}$ FE $_{ 37} \rightarrow \hspace*{0.2cm}^{  63}_{25}$ MN $_{ 38}$}}\\
\hline
Aden & T9& E-Cap &   E+Dec &  NuEn &  GamEn &  NeuEn &  PNEm \\
\hline
     3.0&  1.0 & -51.372 &    -100 & -51.952 & -52.041 & -89.188 &-40.000 \\
     3.0&  3.0 & -17.320 &    -100 & -17.406 & -17.725 & -30.229 &-10.000 \\
     3.0& 30.0 &   1.675 &    -100 &   2.677 &   2.149 &  -0.020 & -2.000 \\
     7.0&  1.0 & -45.553 &    -100 & -46.132 & -46.221 & -83.368 &-40.000 \\
     7.0&  3.0 & -15.605 &    -100 & -15.691 & -16.010 & -28.514 &-10.000 \\
     7.0& 30.0 &   1.678 &    -100 &   2.681 &   2.152 &  -0.017 & -2.000 \\
    11.0&  1.0 &   2.883 &    -100 &   3.859 &   3.160 &   1.490 & -1.000 \\
    11.0&  3.0 &   2.918 &    -100 &   3.882 &   3.278 &   1.703 & -1.000 \\
    11.0& 30.0 &   4.891 &    -100 &   6.097 &   5.364 &   3.617 & -2.000 \\
\hline
\end{tabular}\\
\vspace{0.20cm}\\
\noindent
\begin{tabular}{rr|rrrrrr}
\multicolumn{7}{c}{\normalsize{\hspace*{3.5cm}$^{  63}_{27}$ CO $_{ 36} \rightarrow \hspace*{0.2cm}^{  63}_{28}$ NI $_{ 35}$}}\\
\hline
Aden & T9& E+Cap &   E-Dec &  ANuEn &  GamEn &  NeuEn &  PNEm \\
\hline
     3.0&  1.0 &  -5.841 &  -0.583 &  -0.322 &  -0.885 & -26.117 &-20.000 \\
     3.0&  3.0 &  -2.638 &  -0.409 &  -0.124 &  -0.704 &  -9.550 & -9.000 \\
     3.0& 30.0 &   2.526 &   0.290 &   3.644 &   2.673 &   3.114 & -0.100 \\
     7.0&  1.0 & -11.661 &  -0.684 &  -0.457 &  -1.092 & -29.827 &-30.000 \\
     7.0&  3.0 &  -4.344 &  -0.477 &  -0.219 &  -0.826 & -10.788 &-10.000 \\
     7.0& 30.0 &   2.522 &   0.289 &   3.640 &   2.670 &   3.111 & -0.100 \\
    11.0&  1.0 &    -100 &    -100 &    -100 &    -100 &    -100 &-100\\
    11.0&  3.0 & -42.821 & -35.428 & -35.593 & -36.318 & -47.737 &-10.000 \\
    11.0& 30.0 &  -1.330 &  -2.898 &  -0.214 &  -1.150 &  -0.747 & -0.100 \\
\hline
\end{tabular}\\
\vspace{0.20cm}\\
\begin{tabular}{rr|rrrrrr}
\multicolumn{7}{c}{\normalsize{\hspace*{3.5cm}$^{  63}_{27}$ CO $_{ 36} \rightarrow \hspace*{0.2cm}^{  63}_{26}$ FE $_{ 37}$}}\\
\hline
Aden & T9& E-Cap &   E+Dec &  NuEn &  GamEn &  NeuEn &  PNEm \\
\hline
     3.0&  1.0 & -36.688 &    -100 & -37.263 & -37.701 & -65.532 &-30.000 \\
     3.0&  3.0 & -12.992 &    -100 & -13.069 & -13.606 & -21.387 & -8.000 \\
     3.0& 30.0 &   2.008 &    -100 &   3.014 &   2.158 &   2.223 & -0.200 \\
     7.0&  1.0 & -30.869 &    -100 & -31.444 & -31.881 & -59.712 &-30.000 \\
     7.0&  3.0 & -11.277 &    -100 & -11.354 & -11.891 & -19.671 & -8.000 \\
     7.0& 30.0 &   2.011 &    -100 &   3.018 &   2.162 &   2.227 & -0.200 \\
    11.0&  1.0 &   4.392 &    -100 &   5.486 &   4.630 &   2.580 & -2.000 \\
    11.0&  3.0 &   4.409 &    -100 &   5.506 &   4.651 &   2.606 & -2.000 \\
    11.0& 30.0 &   5.208 &    -100 &   6.417 &   5.315 &   5.461 & -0.100 \\
\hline
\end{tabular}\\
\vspace{0.20cm}\\
\noindent
\begin{tabular}{rr|rrrrrr}
\multicolumn{7}{c}{\normalsize{\hspace*{3.5cm}$^{  63}_{28}$ NI $_{ 35} \rightarrow \hspace*{0.2cm}^{  63}_{29}$ CU $_{ 34}$}}\\
\hline
Aden& T9&  E+Cap &   E-Dec &  ANuEn &  GamEn &  ProEn &  PPEm \\
\hline
     3.0&  1.0 &  -8.724 &  -7.480 &  -8.167 &  -9.198 & -41.326 &-30.000 \\
     3.0&  3.0 &  -5.226 &  -5.881 &  -4.932 &  -5.701 & -14.499 & -9.000 \\
     3.0& 30.0 &   1.550 &  -1.924 &   2.609 &   1.582 &   2.241 & -0.100 \\
     7.0&  1.0 & -14.544 & -10.072 & -10.597 & -11.325 & -45.799 &-30.000 \\
     7.0&  3.0 &  -6.934 &  -6.259 &  -6.263 &  -7.026 & -16.212 &-10.000 \\
     7.0& 30.0 &   1.546 &  -1.925 &   2.606 &   1.579 &   2.238 & -0.100 \\
    11.0&  1.0 &    -100 &    -100 &    -100 &    -100 &    -100 &-100\\
    11.0&  3.0 & -45.412 & -42.751 & -42.922 & -43.752 & -54.689 &-10.000 \\
    11.0& 30.0 &  -2.311 &  -5.274 &  -1.254 &  -2.272 &  -1.624 & -0.100 \\
\hline
\end{tabular}\\
\vspace{0.20cm}\\
\begin{tabular}{rr|rrrrrr}
\multicolumn{7}{c}{\normalsize{\hspace*{3.5cm}$^{  63}_{28}$ NI $_{ 35} \rightarrow \hspace*{0.2cm}^{  63}_{27}$ CO $_{ 36}$}}\\
\hline
Aden & T9& E-Cap &   E+Dec &  NuEn &  GamEn &  NeuEn &  PNEm \\
\hline
     3.0&  1.0 & -25.609 &    -100 & -26.179 & -26.980 & -71.468 &-40.000 \\
     3.0&  3.0 & -10.163 &    -100 & -10.221 & -10.960 & -23.465 &-10.000 \\
     3.0& 30.0 &   1.766 &    -100 &   2.770 &   2.235 &   1.687 & -0.300 \\
     7.0&  1.0 & -19.792 &    -100 & -20.362 & -21.162 & -65.648 &-40.000 \\
     7.0&  3.0 &  -8.452 &    -100 &  -8.510 &  -9.249 & -21.750 &-10.000 \\
     7.0& 30.0 &   1.770 &    -100 &   2.774 &   2.238 &   1.691 & -0.300 \\
    11.0&  1.0 &   4.391 &    -100 &   5.484 &   5.018 &   1.434 & -4.000 \\
    11.0&  3.0 &   4.395 &    -100 &   5.488 &   5.035 &   1.491 & -3.000 \\
    11.0& 30.0 &   4.985 &    -100 &   6.187 &   5.425 &   4.951 & -0.200 \\
\hline
\end{tabular}\\
\vspace{0.20cm}\\
\noindent
\begin{tabular}{rr|rrrrrr}
\multicolumn{7}{c}{\normalsize{\hspace*{3.5cm}$^{  63}_{29}$ CU $_{ 34} \rightarrow \hspace*{0.2cm}^{  63}_{30}$ ZN $_{ 33}$}}\\
\hline
Aden& T9&  E+Cap &   E-Dec &  ANuEn &  GamEn &  ProEn &  PPEm \\
\hline
     3.0&  1.0 & -23.248 & -24.969 & -23.353 & -25.044 & -59.607 &-40.000 \\
     3.0&  3.0 &  -9.382 & -10.934 &  -9.202 & -10.090 & -18.879 & -9.000 \\
     3.0& 30.0 &   1.978 &  -5.886 &   3.020 &   2.394 &   2.137 & -0.200 \\
     7.0&  1.0 & -29.068 & -26.814 & -27.427 & -27.783 & -65.427 &-40.000 \\
     7.0&  3.0 & -11.097 & -11.526 & -10.884 & -11.658 & -20.594 & -9.000 \\
     7.0& 30.0 &   1.974 &  -5.888 &   3.017 &   2.390 &   2.133 & -0.200 \\
    11.0&  1.0 &    -100 &    -100 &    -100 &    -100 &    -100 &-100\\
    11.0&  3.0 & -49.574 & -49.844 & -49.348 & -50.090 & -59.072 & -9.000 \\
    11.0& 30.0 &  -1.885 &  -9.388 &  -0.845 &  -1.468 &  -1.729 & -0.200 \\
\hline
\end{tabular}\\
\vspace{0.20cm}\\
\begin{tabular}{rr|rrrrrr}
\multicolumn{7}{c}{\normalsize{\hspace*{3.5cm}$^{  63}_{29}$ CU $_{ 34} \rightarrow \hspace*{0.2cm}^{  63}_{28}$ NI $_{ 35}$}}\\
\hline
Aden & T9& E-Cap &   E+Dec &  NuEn &  GamEn &  NeuEn &  PNEm \\
\hline
     3.0&  1.0 & -10.477 & -12.991 & -10.783 & -11.428 & -45.929 &-30.000 \\
     3.0&  3.0 &  -5.041 &  -7.378 &  -4.873 &  -5.372 & -15.266 &-10.000 \\
     3.0& 30.0 &   2.399 &  -2.837 &   3.473 &   2.947 &   2.391 & -0.500 \\
     7.0&  1.0 &  -5.864 & -12.991 &  -6.042 &  -6.852 & -40.110 &-30.000 \\
     7.0&  3.0 &  -3.466 &  -7.362 &  -3.281 &  -3.720 & -13.551 &-10.000 \\
     7.0& 30.0 &   2.403 &  -2.836 &   3.477 &   2.951 &   2.394 & -0.500 \\
    11.0&  1.0 &   4.565 & -12.991 &   5.721 &   4.990 &   3.872 & -0.500 \\
    11.0&  3.0 &   4.574 &  -7.361 &   5.732 &   5.007 &   3.916 & -0.500 \\
    11.0& 30.0 &   5.326 &  -2.648 &   6.615 &   5.766 &   5.572 & -0.300 \\
\hline
\end{tabular}\\
\vspace{0.20cm}\\
\noindent
\begin{tabular}{rr|rrrrrr}
\multicolumn{7}{c}{\normalsize{\hspace*{3.5cm}$^{  63}_{30}$ ZN $_{ 33} \rightarrow \hspace*{0.2cm}^{  63}_{31}$ GA $_{ 32}$}}\\
\hline
Aden& T9&  E+Cap &   E-Dec &  ANuEn &  GamEn &  ProEn &  PPEm \\
\hline
     3.0&  1.0 & -33.498 & -43.836 & -33.451 & -35.950 & -53.053 &-20.000 \\
     3.0&  3.0 & -12.352 & -18.952 & -12.191 & -12.944 & -18.017 & -5.000 \\
     3.0& 30.0 &   1.273 &  -8.675 &   2.285 &   0.198 &   2.095 & -0.020 \\
     7.0&  1.0 & -39.318 & -47.138 & -39.270 & -41.770 & -58.873 &-20.000 \\
     7.0&  3.0 & -14.068 & -19.769 & -13.906 & -14.659 & -19.732 & -5.000 \\
     7.0& 30.0 &   1.269 &  -8.676 &   2.282 &   0.194 &   2.091 & -0.020 \\
    11.0&  1.0 &    -100 &    -100 &    -100 &    -100 &    -100 &-100\\
    11.0&  3.0 & -52.545 & -58.183 & -52.384 & -53.137 & -58.209 & -5.000 \\
    11.0& 30.0 &  -2.593 & -12.192 &  -1.580 &  -3.665 &  -1.771 & -0.020 \\
\hline
\end{tabular}%
\\
\vspace{0.20cm}\\

\begin{center}
\textbf{\Large EXPLANATION OF TABLE}
\end{center}
\vspace{0.7in}
\textbf{\large TABLE B. $\mathbf{fp}$-Shell Nuclei Weak Rates of Neutron-Rich and Proton-Rich Nuclei in Stellar Matter}
\vspace{0.4in}
\noindent
\newline
The weak interaction rates for neutron-rich and proton-rich nuclei are calculated using the mass formula of Myers and Swiatecki \cite{Mye96}. The explanation of the table is similar to that of Table~A. The superscript (MS) on each parent nucleus indicates that the mass formula \cite{Mye96} is used for the calculation of Q-values and separation energies.
\clearpage
\sffamily \normalsize
\noindent
\\
\vspace{0.20cm}\\

\end{document}